\documentclass[]{aastex631}
\usepackage{graphicx}

\usepackage[singlelinecheck=false]{}

\newcommand{\oiii}{\,\hbox{[\ion{O}{3}]}}
\newcommand{\arii}{\,\hbox{[\ion{Ar}{2}]}}
\newcommand{\ariii}{\,\hbox{[\ion{Ar}{3}]}}

\newcommand{\arv}{\,\hbox{[\ion{Ar}{5}]}}
\newcommand{\mgv}{\,\hbox{[\ion{Mg}{5}]}}
\newcommand{\neii}{\,\hbox{[\ion{Ne}{2}]}}
\newcommand{\neiii}{\,\hbox{[\ion{Ne}{3}]}}
\newcommand{\feii}{\,\hbox{[\ion{Fe}{2}]}}
\newcommand{\fevii}{\,\hbox{[\ion{Fe}{7}]}}

\newcommand{\siv}{\,\hbox{[\ion{S}{4}]}}
\newcommand{\nev}{\,\hbox{[\ion{Ne}{5}]}}
\newcommand{\nevi}{\,\hbox{[\ion{Ne}{6}]}}

\newcommand{\msun}{\,\hbox{$M_{\odot}$}}

\newcommand{\kms}{\,\hbox{\hbox{km}\,\hbox{s}$^{-1}$}}
\newcommand{\degree}{\ensuremath{^\circ}}

\newcommand{\htwo}{\,\hbox{$\rm{H_ 2}$}}

\newcommand{\ang}{\,\hbox{\AA}}

\received{June 3, 2024}
\accepted{October 21, 2024}

\shorttitle{Jet-driven outflows, bow shocks, and high excitation of the gas in the MIRI data of IC5063}
\shortauthors{Dasyra, K. M., et al.}

\begin{document}

\title{ A case study of gas impacted by black-hole jets with the JWST: \\ outflows, bow shocks, and high excitation of the gas in the galaxy IC5063}

\author[0000-0002-1482-2203]{Kalliopi M. Dasyra}
\affiliation{Department of Astrophysics, Astronomy \& Mechanics, Faculty of Physics, National and Kapodistrian University of Athens, Panepistimioupolis Zografou, 15784 Athens, Greece}
\author[0000-0001-6757-3098]{Georgios F. Paraschos}
\affiliation{Max-Planck-Institut für Radioastronomie, Auf dem Hügel 69, D-53121 Bonn, Germany}
\author[0000-0003-2658-7893]{Francoise Combes}
\affiliation{Observatoire de Paris, LERMA, Coll\`ege de France,
CNRS, PSL University, Sorbonne University, Paris, France}
\author[0000-0001-8718-3732]{Polychronis Patapis}
\affiliation{Institute of Particle Physics and Astrophysics, ETH Zurich, Wolfgang-Pauli-Str 27, 8093, Zurich, Switzerland}
\author[0000-0003-3367-3415]{George Helou}
\affiliation{Caltech/IPAC, 1200 E. California Boulevard, Pasadena, CA 91125, USA}
\author[0000-0001-5650-4206]{Michalis Papachristou}
\affiliation{Department of Astrophysics, Astronomy \& Mechanics, Faculty of Physics, National and Kapodistrian University of Athens, Panepistimioupolis Zografou, 15784 Athens, Greece}
\author[0000-0001-9490-899X]{Juan-Antonio Fernandez-Ontiveros}
\affiliation{Centro de Estudios de Física del Cosmos de Aragón (CEFCA), Plaza San Juan 1, E-44001, Teruel, Spain}
\author[0000-0003-2733-4580]{Thomas G. Bisbas}
\affiliation{Research Center for Astronomical Computing, Zhejiang Lab, Hangzhou, 311100, China}
\author[0000-0001-8840-1551]{Luigi Spinoglio}
\affiliation{Istituto di Astrofisica e Planetologia Spaziali (INAF–IAPS), Via Fosso del Cavaliere 100, I-00133 Roma, Italy}
\author[0000-0003-3498-2973]{Lee Armus}
\affiliation{California Institute of Technology}
\author[0000-0001-6919-1237]{Matthew Malkan}
\affiliation{Department of Physics and Astronomy, UCLA, Los Angeles, CA 90095-1547, USA}
\begin{abstract}

We present James Webb Space Telescope MIRI data of the inner $\sim$3$\times$2kpc$^2$ of the galaxy IC5063, in which the jets of a supermassive black hole interact with the gaseous disk they are crossing. Jet-driven outflows were known to be initiated along or near the jet path and to modify the stability of molecular clouds, possibly altering their star formation properties. The MIRI data, of unprecedented resolution and sensitivity in the infrared, now reveal that there are more than ten discrete regions with outflows, nearly doubling the number of such known regions. Outflows exist near the radio lobes, at the nucleus, in a biconical structure perpendicular to the jet, and in a bubble moving against the disk. In some of them, velocities above escape velocity are observed. Stratification is also observed, with higher ionization or excitation gas attaining higher velocities. More outflows and bow shocks, found further away from the nucleus than the radio lobes, in regions without significant radio emission, reveal the existence of past or weak radio jets that interacted with the interstellar medium. The coincidence of the bow shocks with the optical extended emission line region (EELR) suggests that the jets also contributed to the gas ionization. Maps of the \htwo\ gas excitation temperature, T$_{ex}$, indicate that the molecular gas is most excited in regions with radio emission. There, T$_{ex}$ is $>$100\,K higher than in the EELR interior. We argue that a combination of jet-related shocks and cosmic rays is likely responsible for this excess molecular gas excitation. 

\end{abstract}

\keywords{Interstellar medium  --- Molecular gas  --- Galaxy winds  --- Radio jets ---  Galaxies }

\section{Introduction} 
\label{sec:intro}

The advent of the James Webb Space Telescope (JWST), with its unprecedented spatial resolution and sensitivity in the infrared, enabled the astronomical community to make fast advancements in the field of galaxy evolution, by looking further in time for distant galaxies and deeper into nearby galaxies. Besides the detection of the first galaxies and even the first stars in them (e.g., \citealt{atek22,castellano22,naidu22,finkelstein22,finkelstein23,donnan23a,donnan23b,harikane23a,leung23,maiolino23a,yan23}), a major galaxy evolution theme to be revised by the JWST is the impact of black holes (BHs) on galaxy evolution (including via outflows; \citealt{armus23,rupke23,cresci23,vayner24,leftley24}). Cycle 1 surveys already indicated that many more active galactic nuclei (AGN) exist early, above redshift $z$$>$3-4 \citep{barro23,juodzbalis23,kocevski23,labbe23,lyu23}, compared to what was known from past observations in X-rays or other wavelengths,  and that some black holes grew too fast too soon compared to the local BH mass vs. host galaxy relation (e.g., \citealt{harikane23b,larson23,maiolino23b} for 4$<$z$<$8.5). 

If black holes started accumulating their mass earlier, it is likely that their radio-mode feedback phases, operating via radio jets at low Eddington accretion rates, could have started earlier. The radio-mode feedback, statistically operating after the  radiation-mode feedback in simulations, has long been deemed critical for the reproduction of the observed number of massive galaxies in the local Universe (e.g., \citealt{bower06,croton06}). Therefore, BH jets could have propagated earlier, through younger galaxies with more massive interstellar medium (ISM) reservoirs. Follow up of radio catalogues with the JWST and the Atacama Large Millimeter Array (ALMA) indeed led to the discovery of new radio galaxies in the early Universe. For example, in the COSMOS field alone, \citet{endsley22} and \citet{labrides24} detected radio galaxies at z$\sim$6.8 and z$\sim$7.7, respectively. Both of them have intense star formation. Similarly, one of the major results of a recent ALMA survey of radio galaxies \citep{audibert22} was that the gas reservoirs of radio galaxies at 1$<$z$<$3, when detected, are comparable to the gas reservoirs of normal galaxies at the same redshift. This result has important implications for galaxy evolution, because radio jets have a tremendous potential to disturb star formation when propagating through gas-rich hosts. Hydrodynamical simulations have shown that when relativistic jets pass through a gaseous clumpy disk, they contribute to the disk dissipation by greatly dispersing dense clouds (depending on the ISM geometry, clumpiness, and density) and by forming a mass loaded wind that entrains even more material \citep{wagner11,wagner13,gaibler12,mukherjee18}. 

The turmoil that jets cause as they propagate through a dense ISM is demonstrated in this paper, with the aid of cycle 1 JWST data of the nearby, early-type galaxy IC5063 (taken for the program 2004; PI Dasyra). IC5063 is well suited for demonstrating the capabilities of the JWST, as it is a galaxy known for its numerous jet-cloud interactions. These interactions occur as radio jets cross a disk of dense gas for hundreds of parsecs, along a trajectory nearly parallel to the disk plane, prior to escaping it \citep{morganti98}. The jets collide with dense ISM clouds at the two radio lobes, $\sim$300-500 pc away from the nucleus, driving outflows. An outflow has been seen in absorption in HI in front of one of the radio lobes \citep{oosterloo00,morganti07}. Outflows have also been seen in emission in several regions, in tracers of all gas phases - atomic \citep{morganti07, dasyra15, congiu17,venturi21} and molecular \citep{tadhunter14,morganti15,dasyra15,dasyra16,dasyra22}.  In Very Large Telescope (VLT) SINFONI data, \citet{dasyra15} found numerous outflows starting at or near the jet trail: at least 6 discrete outflow starting points were reported for the central 2$\times$2kpc$^2$ of this galaxy, including at the two radio lobes. In some of these locations, gas moving faster than the escape velocity was detected. Moreover, the gas in nearly all of the inner 1.5 square kiloparsec of the galaxy shows distorted kinematic signatures due to the jet. In VLT MUSE data, \citet{venturi21} showed that, in the ionized gas as seen by \oiii, the turbulence in the galaxy may reach $\sim$4\,kpc in both directions perpendicular to the jet. \citet{morganti15} proposed even the propagation of a homonegeous jet-inflated cavity (a.k.a. cocoon) in the inner region of IC5063 based on the low rotational number J CO data distribution. The mass of all outflowing gas is at least 6$\times$10$^6$ \msun ,  when taking into account the mass of the neutral gas and the cold molecular gas, using their most-abundant gas tracers, HI and CO \citep{morganti07,dasyra16}. In this computation, the cold \htwo\ gas mass was determined using a lower than Galactic CO luminosity to \htwo\ mass factor (\citealt{dasyra16};  see also \citealt{bolatto13}), because some of the molecular gas emission in the outflow's location was found to be optically thin \citep{dasyra16,oosterloo17}, potentially related to cloud evaporation in the flow.

IC5063 has a large and rich set of available ancillary data, which allow novel experiments on the impact of jets on star formation to be carried out. In a recent paper, \citet{dasyra22} showed that ongoing star formation (SF) changes linked to the jet can be sought via stability analysis of jet-impacted clouds. The concept behind that paper was simple: whether a cloud will collapse, and contribute to the enhancement of SF, or disperse, and add to the suppression of SF, depends upon the interplay of its self gravity with changes in its internal and external pressure. The pressure measurement can be challenging, yet it is feasible. It requires radiative transfer modeling of the line emission, if the latter is assumed not to be in local thermodynamic equilibirum (LTE). For the cold and dense molecular gas clouds, \citet{dasyra22} derived the internal pressure P$_{i}$  by fitting the CO (1-0) up to (4-3) and HCO+ (4-3)/(1-0) ALMA data with the astrochemical and radiative transfer code (3D-PDR; \citealt{bisbas12}), taking into account various heating sources: UV radiation, jet-related cosmic rays, and mechanical heating (associated with shocks). They fitted 3D-PDR models to each pixel of the spatially resolved ALMA data and found that $P_{i}$
increases from 5$\times$10$^4$ K cm$^{-3}$ (in ambient clouds a few hundred pc away from the jet trail) to 10$^6$K cm$^{-3}$ (in jet-impacted clouds at the radio lobes). For the external pressure of the molecular clouds P$_{e}$, they considered it to be the pressure of the surrounding ionized medium, which was derived from standard diagnostics of the ionized gas temperature ([\ion{N}{2}]) and density ([\ion{S}{2}]) in MUSE data that are applicable for Boltzmann populations under LTE. Similarly, they found an increase of more than one order of magnitude at the radio lobes for P$_{e}$. They also found that P$_{e}$ frequently exceeds P$_{i}$, which led them to conclude that we are observing the expansion of an overpressurized ionized gas bubble, perhaps the jet cocoon, that engulfs and compresses clouds as it propagates. Simultaneously, in other, underpressurized regions,  plugging P$_{i}$ and P$_{e}$ in a modified virial equation that is appropriate for molecular clouds indicated that some clouds can become gravitationally unbound and lose outer layers. This stability analysis therefore indicated that both star formation induction and suppression can occur in IC5063 \citep{dasyra22}.

We obtained new JWST data with the goals of studying the molecular and ionized gas distribution, excitation, and kinematics, finding the extent and mass of jet-driven outflows, comparing the loci of these outflows in unobscured wavelengths with the loci of over/under-pressurized regions,
and performing another calculation of the molecular gas pressure at gas layers directly probed by the warm \htwo. Many of these goals are addressed in this paper, whereas the gas pressure will be addressed, with the use of detailed shock modeling, in a future paper.

\section{The Data}
\label{sec:data}

\begin{figure}[ht!]
\plotone{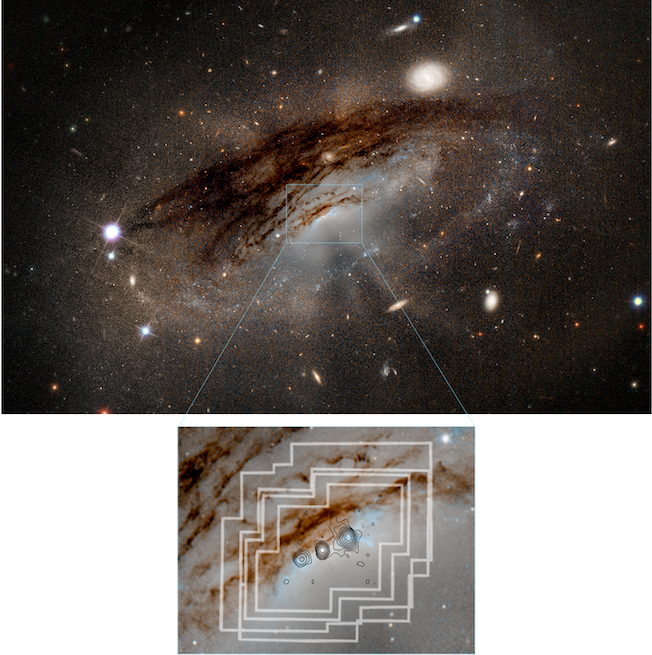}
\caption{ IC5063 as seen in multiple optical bands, from data of two Hubble Space Telescope (HST) snapshot proposals (15446 and 15444) and the Legacy Survey, DR9 release. Image credit: NASA/ESA/CSA/ J. Schmidt, attribution: NASA/ESA/ Aaron Barth / Julianne Dalcanton / DECaM Legacy Survey / Judy Schmidt, license: CC BY 2.0. For the Legacy Survey data, DECam g is shown in blue, DECam r is in green and DECam z is in red. For the HST data, ACS/WFC F606W is in blue and ACS/WFC F814W is in red (see also \citealt{maksym20} for the HST data). In this image, north is 3.29° clockwise from up. The inset shows the mosaic coverage in the four channels of the MIRI MRS observations for the cycle 1 program 2004 (PI Dasyra). The contours show the radio emission (including the radio lobes) of the galaxy at 17.8 GHz \citep{morganti07}.  \label{fig:FOV}}
\end{figure}

We acquired data of IC5063 with the MIRI instrument onboard the JWST  for the cycle 1 program 2004 (PI: Dasyra). MIRI was used in its medium resolution spectrometer (MRS) integral-field-unit (IFU) mode \citep{wells15,argyriou23}. The observations were executed the 13th of May 2023 and targeted the inner 13\arcsec $\times$8\arcsec\ of the galaxy in ch1, and up to the inner 19\arcsec $\times$14\arcsec\ in ch4, in an area of the nucleus of IC5063 that follows the radio jet trail (Fig.~\ref{fig:FOV}). A mosaic of four different tiles was used to cover the area of interest, because the MIRI IFU field of view (FOV), is  3.3\arcsec $\times$3.7\arcsec\ wide in channel 1 and 6.6\arcsec $\times$7.7\arcsec\ wide in channel 4. Additionally, a four-dither pattern was used for each tile, to assist with drizzling and with the identification and elimination of problematic pixels. In each pointing position, the exposure was either executed in a single up-the-ramp integration or split between two up-the-ramp integrations, following the wavelength-dependent saturation indications of the Exposure Time Calculator (ETC). One up-the ramp integration per exposure was used for the short detector (channel 1 and 2) data, and two up-the ramp integrations were chosen for most long detector (channel 3 and 4) data. Separate sky observations for the background elimination were also carried out. They had an identical exposure time and the same number of integrations per exposure to that of the science observations, for a single mosaic tile and a four-dither pattern.

The level 1 data were downloaded with the "astroquery" package of python, which queries the Mikulski Archive for Space Telescopes (MAST)\footnote{The pipeline-version of our data can be retrieved from MAST using the link \dataset[https://doi.org/10.17909/jvw9-5x80]{https://doi.org/10.17909/jvw9-5x80}.}. The data were reduced with version 1.12.5 of the JWST (Python-based) pipeline\footnote{This pipeline version can be retrieved from the link https://doi.org/10.5281/zenodo.10022973 .} and the jwst\_1202.pmap content of the Calibration Reference Data System (CRDS). In the first stage of the pipeline, the function "Detector1" was run, which primarily corrects for detector artifacts and produces detection rates from time-dependent slopes. All parameters were set to their default values for exposures with two integrations. Then, the values of pixels with cosmic ray (CR) hits in one integration were replaced by their values in the other integration. For exposures with a single integration, "Detector1" was run with the additional use of the "find\_showers" parameter, so that CRs could be identified. Their flux was modeled and subtracted from the relevant data by the pipeline. We performed two extra reduction steps of our own to improve the data products. The first step was the subtraction of the background from the science frames. A master background was created for each band (channel and sub-band) as the median of the four available exposures. It was then subtracted from each science frame of the same band. This two-dimensional background subtraction had the optimal result with respect to the removal of artifacts compared to the one-dimensional or three-dimensional subtraction that is offered at later stages of the pipeline. Our second additional step was the masking of bad pixels in the science data using a master bad pixel mask per band. We created this mask by merging the CRDS mask file with a mask of outliers ($>$5$\sigma$) in the master sky frame. In the background-subtracted science data, the replacement values of bad pixels were computed by interpolation of their neighbors' values using the "Gaussian2DKernel" function of the "astropy" package and a standard deviation of one pixel.

\begin{figure}[ht!]
\centering
\includegraphics[width=1.1\textwidth, trim=4cm 0cm 0cm 0cm, clip]{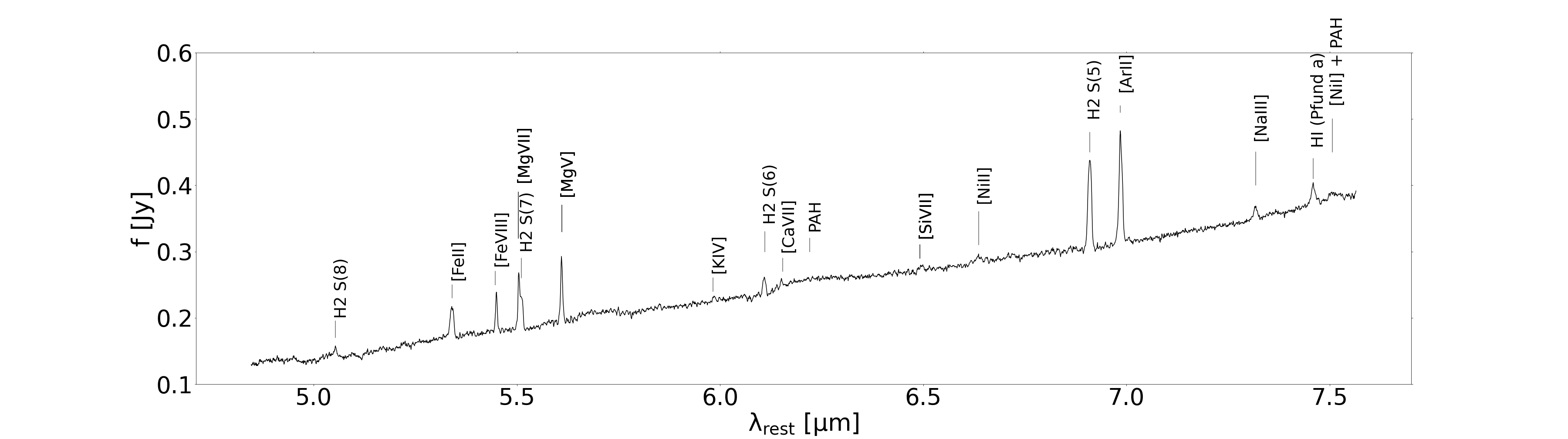}\\
\includegraphics[width=1.1\textwidth, trim=4cm 0cm 0cm 0cm, clip]{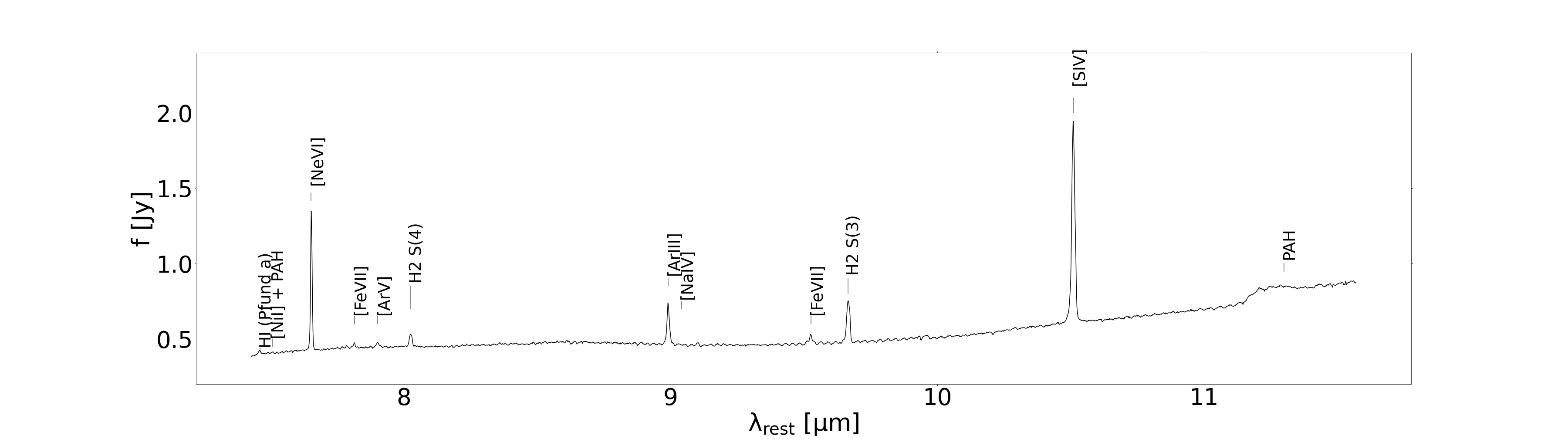}\\
\includegraphics[width=1.1\textwidth, trim=4cm 0cm 0cm 0cm, clip]{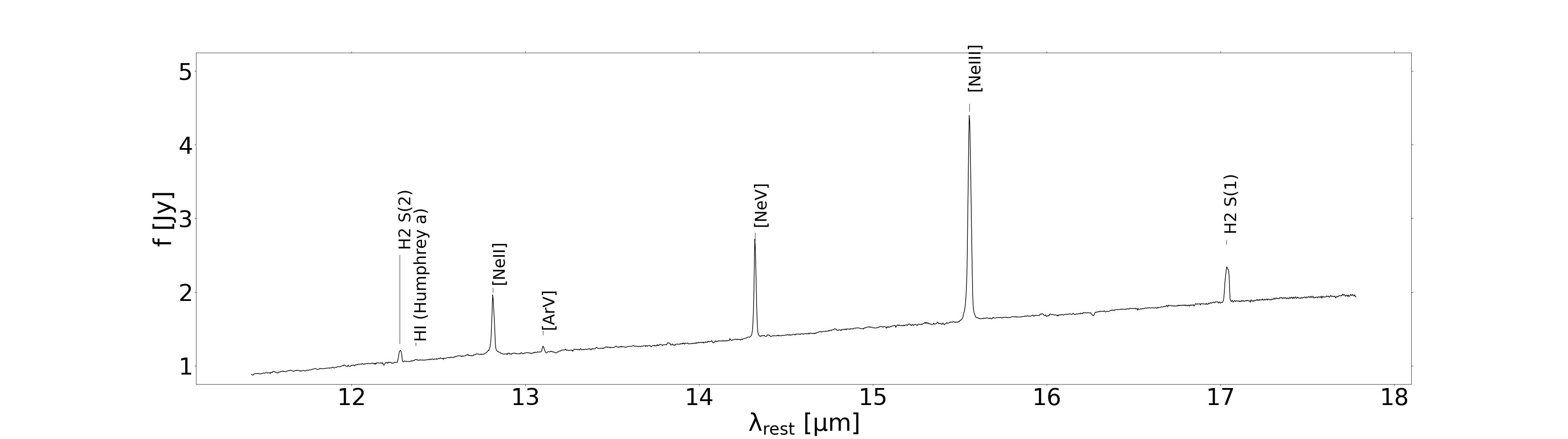}\\
\includegraphics[width=1.1\textwidth, trim=4cm 0cm 0cm 0cm, clip]{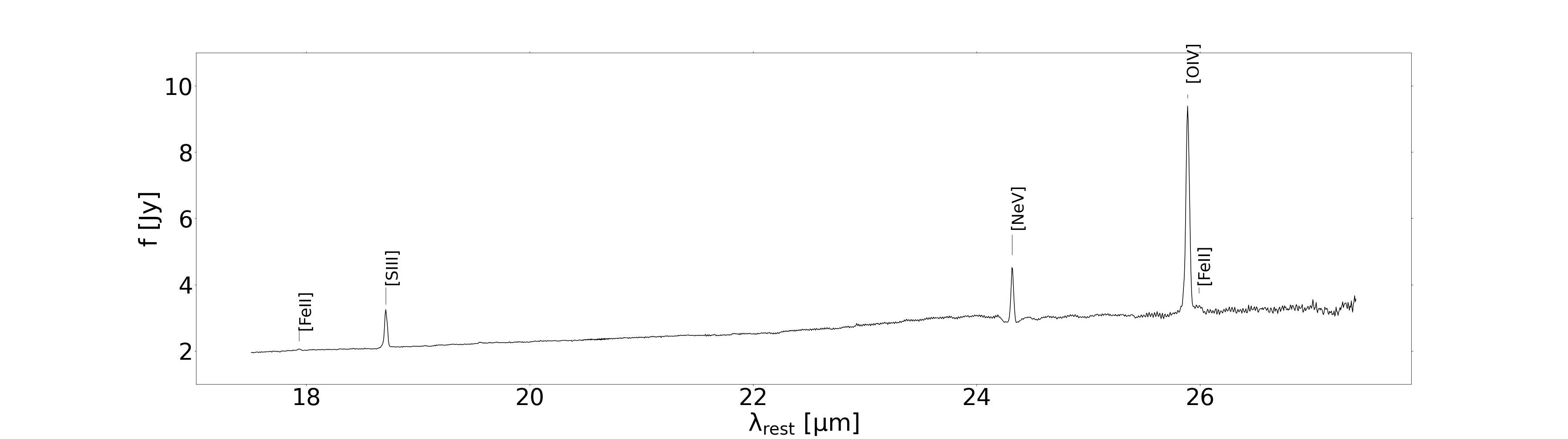}\\
\caption{ Integrated spectrum in the entire mosaic of channel 1 through channel 4 (from top to bottom). 
\label{fig:spectrum}}
\end{figure}

We input the background-subtracted data to the second stage of the pipeline, "Spec2Pipeline", for all relevant calibrations prior to the building of the 3-dimensional cube. These include the  world coordinate system (wcs) information buildup, the flatfielding, the flux calibration, the straylight correction, the fringe removal, and the rectification of the data. Besides all functions that run by default, we also enabled the residual fringe removal function (and disabled the background subtraction function). During this step, we also created a version of the cubes without the straylight correction, to check for the impact of the cruciform artifact (see Section~
\ref{Appendix:cruciform}).

Calibrated cubes were created, which were then processed by "Spec3Pipeline", for the final stage of the pipeline reduction, which merges individual exposure and pointing position cubes into mosaics of wcs-aligned cubes per band or channel. We opted to produce mosaics per band, for a custom handling of data in overlapping spectral regions. The outlier detection and the default parameter values (e.g., drizzle for weighing) were used. Further 1-dimensional residual fringe correction was applied. For the extraction of the spectrum of the nuclear point source, autocentering of the aperture location was turned on. No further background subtraction was used, so all pertinent functions were skipped (e.g., "mrs\_imatch"). The footprint of the created mosaics, which increases with wavelength, is shown in Fig.~\ref{fig:FOV}. Additionally, we converted the units of the final mosaic cubes from MJy/sr to Jy, taking into account the channel-dependent pixel size.

 The integrated spectrum of all the emission in the MIRI mosaic of each band is shown in Fig.~\ref{fig:spectrum}. An impressive number (39) of spectral lines is detected, most of which have spatially-resolved emission (see Section \ref{sec:results-basic}). Unlike the spectral lines, the mid-infrared continuum emission of IC5063 is nearly unresolved, originating in its largest fraction from the nuclear point-spread function (PSF; as seen in ~\ref{Appendix:cont}). Due to flatfielding uncertainties in the long band of channel 4, its flux is on average lower (by $\sim$8\%) than that indicated by the flux in previous channels. This calibration uncertainty also applies to fluxes of the lines in the same channel and band. It is comparable to the nominal flux calibration accuracy of the instrument (10\%).

Continuum-free cubes have been created with own Python routines for each spectral line using a local continuum image. For this purpose, a continuum image was first created from cube slices of emission-line-free wavelengths, indicatively 3000-4000 \kms\ away from the spectral line rest-frame wavelength. Then, this image was scaled in integrated flux to match the continuum flux per wavelength, according to a first degree polynomial that best fits the continuum emission in the spatially integrated spectrum (Fig.~\ref{fig:spectrum}). Because of the strength of the continuum, the residual noise inside the nuclear PSF that appears after the continuum subtraction can be non-negligible compared to the emission of several spectral lines. This means that the nuclear line fluxes are more uncertain that the line fluxes at other mosaic locations. To visually enhance the appearance of the presented line cubes and images in this paper, we created smoothed cubes, convolved by a Gaussian kernel of 1 pixel (for channels 1,2) or 1.5 pixels (for channels 3,4). We then replaced the pixel values of the original-resolution cubes inside the PSF (only) with those of the smoothened cubes.

The lines were found to have an average z=0.01121. For an adopted $\Lambda$CDM cosmology, with H$_0$=70 \kms\  Mpc$^{-1}$, $\Omega_{M}$=0.3 and $\Omega_{\Lambda}$=0.7, the galaxy has a luminosity distance of 48.4 Mpc. An arcsecond in the sky corresponds to 230 pc in the galaxy at that distance.

\begin{figure}[ht!]
\centering
\includegraphics[width=\textwidth]{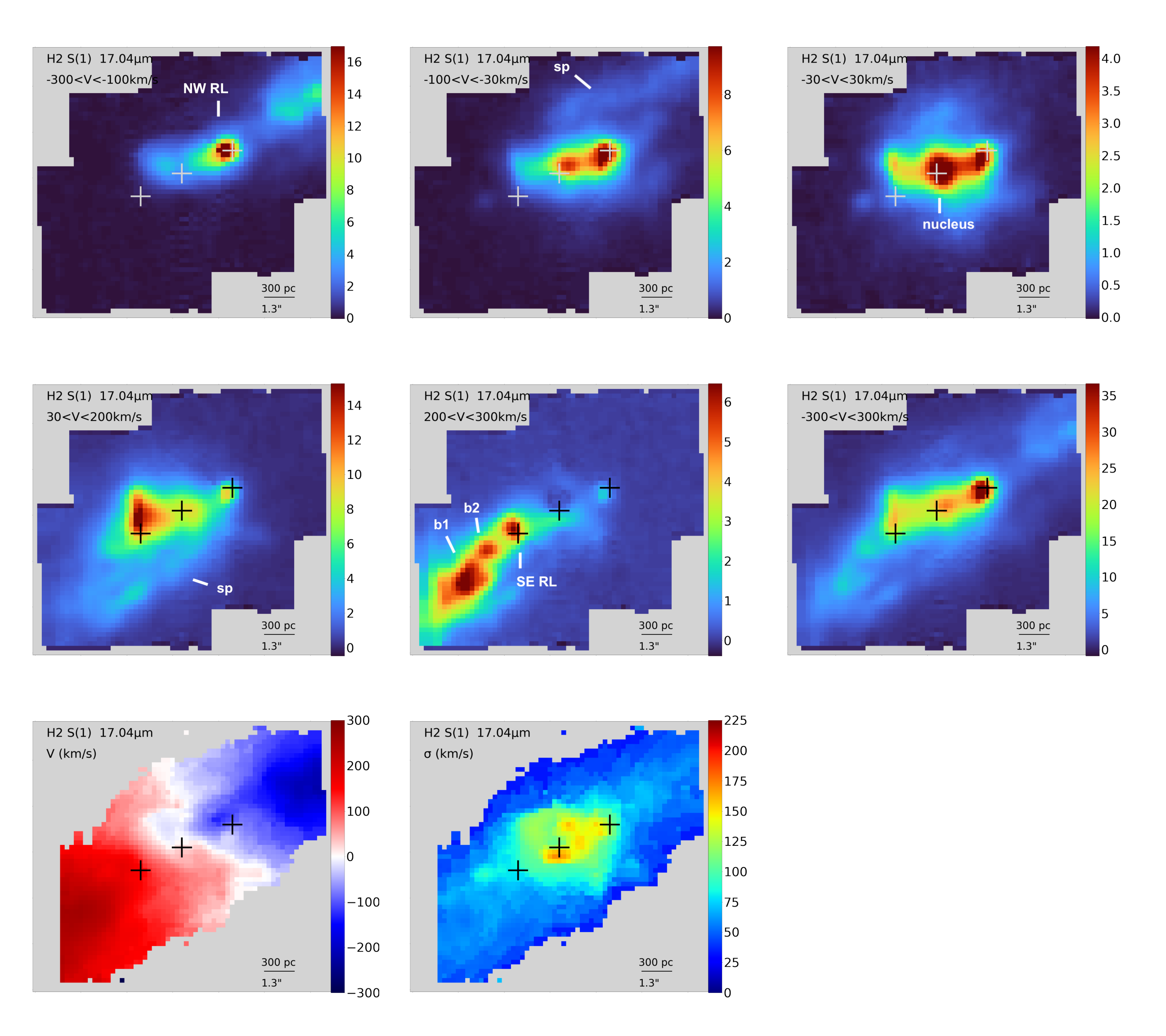}
\caption{Properties of the disk, as best seen in \htwo\ S(1). {\it Upper and middle rows:} Images of the emission at different velocity ranges showing the rotation and structures of interest. {\it Lower row:} Moment 1 and 2 (velocity and velocity dispersion) maps.  North is up. Crosses mark the location of the nucleus and the EELR bases, NW and SE of the nucleus, as derived from the ionized gas emission at rest-frame velocity in all MIRI data (see the Appendix). Following \citet{morganti98}, the EELR bases are considered to be the two radio lobes (see also Fig.~\ref{fig:FOV}). Spiral arms are marked as `sp', blobs are marked with `b', and radio lobes with `RL'. 
\label{fig:disk}
}
\end{figure}

\section{Results}
\label{sec:results}

\subsection{Overall emission properties and molecular disk properties}
\label{sec:results-basic}

The continuum emission is largely unresolved, with more than 96\% of its flux in all bands originating from the supermassive black hole vicinity,  at 20:52:02.36 $-$57:04:07.54, according to the location of the PSF in ch1. Given that target acquisition was performed on a nearby star (at 20:52:09.6647 $-$57:04:15.41), the uncertainty in these coordinates is below 0.1$\arcsec$: according to the JWST documentation\footnote{ https://jwst-docs.stsci.edu/}, the telescope's pointing accuracy with target acquisition is better than 0.1$\arcsec$ and potentially as good as 0.014$\arcsec$. The continuum  flux is 0.22($\pm$0.01), 0.95($\pm$0.02), 1.28($\pm$0.02), and 3.0($\pm$0.3) Jy at 6, 12, 14, and 25 $\mu m$ respectively. In the MIRI wavelengths, we primarily see the AGN-heated dust, which is likely located in the black hole torus. By fitting the nuclear emission at 2.2 and 3.8 $\mu m$, \citet{kulkarni98} deduced an indicative torus temperature of $\sim$720\,K. By modeling the torus emission in lower spatial resolution data, \citet{esparza19} provided a torus mass range of $\sim$10$^3$-10$^6$ \msun . The torus clumpiness and compactness have been considered responsible for the detection of crepuscular rays in this galaxy \citep{maksym20}. Its minimum extent, from predictions of the dust sublimation radius is 0.07\,pc  \citep{hoenig10}. Its maximum extent is limited, from the PSF our ch1 data, to $\sim$50\,pc, moderately improving our knowledge from ground-based mid-infrared sub-arcsecond observations (82 pc; \citealt{asmus14}).  A small contribution from the radio core to the nuclear spectrum is also possible, given that some continuum emission (up to 4\%) exists between the nucleus and the radio lobes, which can be attributed to the jet-related synchrotron. Some weak silicate absorption is also seen at 9.7 $\mu m$, with optical depth $\tau_{9.7}$=0.18, which indicates the existence of compact dust in or around the torus (and the radio core). The diffuse dust emission in the MIRI FOV is very weak compared to the nuclear point-source emission. Still, the spatially-resolved PAH emission, in particular that of the brightest 11.3$\mu m$ complex, confirms its presence.
 
\begin{figure}[ht!]
\centering
\includegraphics[width=\textwidth, trim=0cm 2cm 0cm 5cm, clip]{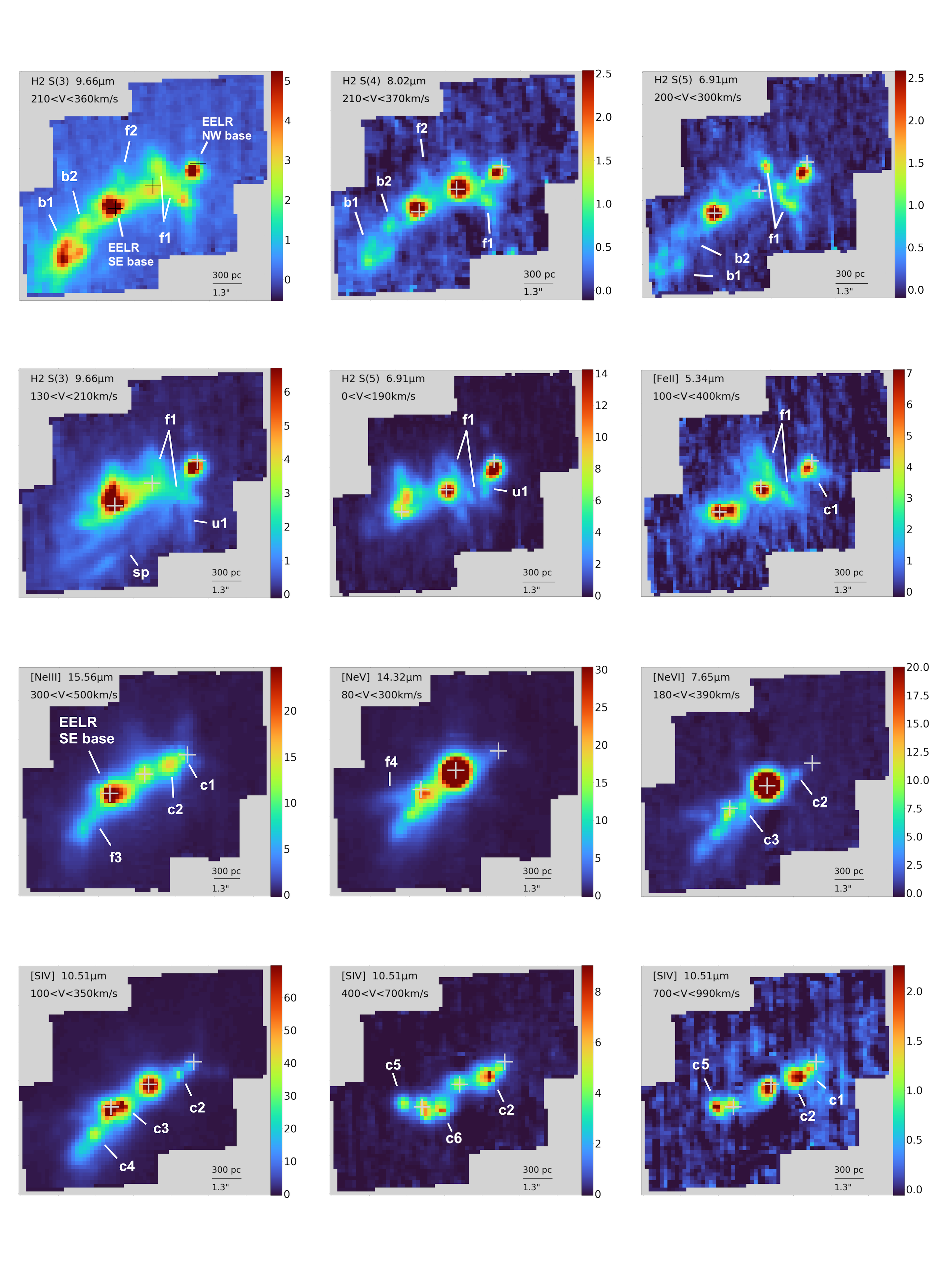}
\caption{ Structures of interest including bow shocks, filaments, discrete outflows,  and structures moving against the disk at positive velocities. Clouds, filaments, bow shocks/bubbles, and spiral arms are respectively marked with `c',`f', `b', and `sp'. The base of the eastern EELR cone is defined by the f3 and f4 filaments. The filament marked with `u1' is of uncertain origin: it is either real or due to an artifact described in the Appendix. 
\label{fig:all_south}}
\end{figure}

\begin{figure}[ht!]
\centering
\includegraphics[width=\textwidth, trim=0cm 2cm 0cm 5cm, clip]{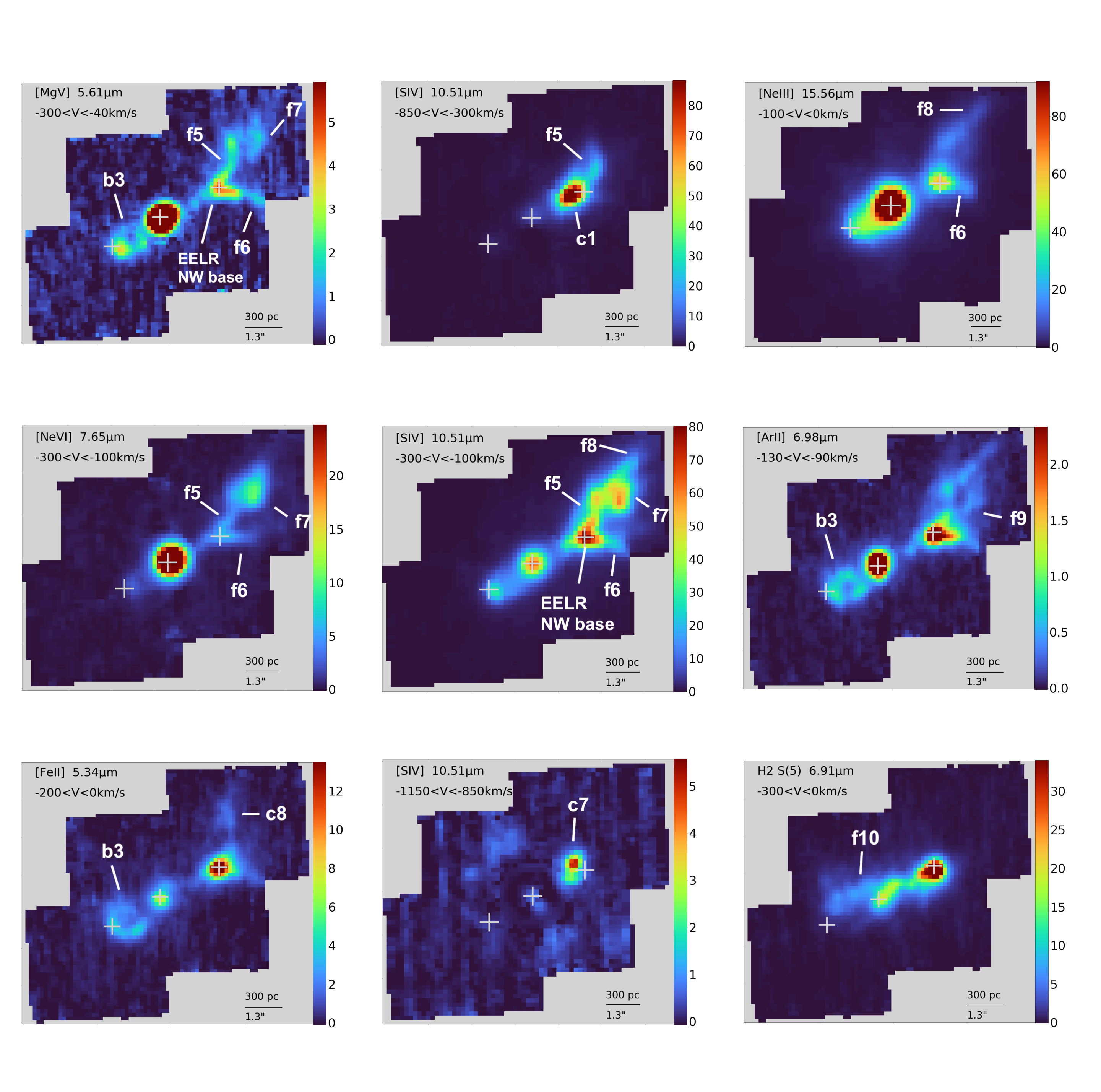}
\caption{ Structures of interest including bubbles, filaments, discrete outflows, and structures moving against the disk at negative velocities. The nomenclature is as in the previous Figure. The base of the western EELR cone is defined by the f5 and f6 filaments.
\label{fig:all_north}}
\end{figure}

In contrast to the continuum emission, most of the line emission is spatially resolved. Thanks to the unprecedented quality of the MIRI data, 
39 lines of molecular and atomic gas phase were detected, and spectral cubes were created for all of them except for the polycyclic aromaric hydrocarbons (PAHs). Images of the integrated emission of all lines are shown in the Appendix. Their total line fluxes are presented in Table~1. The comparison with the previous {\it Spitzer} data is impressive. Less than half of the Table~1 spectral lines were previously present in spatially-integrated {\it Spitzer} Infra-Red Spectrograph (IRS) spectra \citep{tommasin10,panuzzo11, rampazzo13}, even though the IRS field of view was comparable to that of our MIRI mosaic for the short-high (SH) and long-high (LH) modes, and greater than that of our MIRI mosaic for the short-low (SL) and long-low (LL) modes. A bare minimum of kinematic information (i.e., a linewidth) was extracted for a handful of lines in the SH and LH modes \citep{guillard12}. Detailed kinematical analysis is now possible for nearly all lines, revealing out-of-dynamical-equilibrium components in many of them, e.g., in all of the previously detected or undetected \htwo\ lines -  from S(1) to S(8). Depending on the line under examination, the emission probes the disk and regions within it or near it that are excited by the jet, or even the  nucleus.  Images indicative of all features of interest are shown in Figs.~\ref{fig:disk},~\ref{fig:all_north}, and ~\ref{fig:all_south}, for selected velocity ranges.

The spectral line that best probes the disk extent and properties is \htwo\ S(1) (Fig. ~\ref{fig:disk}). Faint spiral arms can be seen in all \htwo\ data from S(1) to S(3). In the upper part of the disk, a spiral arm starting north-east of the nucleus and heading north-west of it, can be seen in all upper panels of Fig.~\ref{fig:disk}. In the lower part of the disk, a fainter spiral structure extending from south-west of the nucleus to south-east of it can be seen in Fig.~\ref{fig:disk} for molecular gas velocities in the range 30$<$V$<$200\kms\ range. Spiral structure with similar orientation has been seen before in CO interferometric data \citep{dasyra16,dasyra22}. Given the detected disk rotation pattern that has been computed as the first moment of the \htwo\ S(1) cube (Fig.~\ref{fig:disk}), then the upper side of the disk is closer to us if the spiral arms are trailing. The detected dust lane location, in the north part of the galaxy (e.g., Fig.~\ref{fig:FOV}), confirms this and verifies that the spiral arms are trailing. Still, the brightest \htwo\ emission does not originate from the spiral arms, but from several clumps along the radio jet axis. In the line connecting the north west (NW) to the south east (SE) radio lobe, the brightest clumps are seen near the radio lobes themselves for all \htwo\ lines, from S(1) to S(8). Other bright clumps also exist further away from the nucleus, in particular for the lower rotational number lines. For example, in the south-east, two bright discrete clumps are seen in \htwo\ S(1) $\sim$ 900 and 1200 pc away from the nucleus, for positive disk velocities (identified as blobs b1 and b2 in Figs.~\ref{fig:disk} and ~\ref{fig:all_south}).

In contrast to the molecular gas, the ionized gas lines do not show any spiral structure at disk velocities (see the Appendix). The emission of most these lines largely follows the jet direction from the nucleus all the way to the radio lobes (see examples in Figs.~\ref{fig:all_south} and ~\ref{fig:all_north}, which separately present images for positive and negative velocity ranges so that regular and irregular kinematics can be easily distinguished). At the radio lobes, the bright emission is attributed to the intense jet cloud-interactions are known to be taking place \citep[e.g.,][]{oosterloo00,morganti07, tadhunter14,dasyra15}. The two radio lobes coincide with the bases of the two cones of the extended emission line region (EELR) \citep{morganti98}, which makes the galaxy appear roughly X-shaped in its large-scale ionized gas distribution in optical wavelengths \citep{colina91}. In the MIRI data, the brightest locations of the ionized gas emission are indeed the two the EELR bases (or radio lobes) and the nucleus. Lines of higher ionizational potential (IP) tend to have more nucleated emission than lines with low IP. For all lines with IPs $>$100 eV (shown after \fevii\ in the first figure of the Appendix), more than two thirds of the emission is circumnuclear, linked to the AGN and to the radio core. From the radio lobes outwards, the emission splits up in two main branches (see, e.g., \mgv\ and \fevii\ in the same figure, for the two branches north-west of the NW radio lobe and \siv\ and \nevi\ for the branches south-east of the SE radio lobe), which is why we alternatively call the EELR bases as bifurcation points. Further out, these branches constitute the edges of the EELR, which is long known to be oriented at a position angle similar to that of the disk instead of perpendicular to it \citep{colina91}. Just like in the optical, the EELR orientation is similar to that of the jet in the mid-infrared. At distances greater than the bifurcation points, all of the ionized gas emission is constrained in the interior of the EELR. Good tracers of the jet-ionized gas along the jet trail and along the EELR edges are \mgv\ and \nevi , thanks to their high IPs (109.3 and 126.2 eV, respectively) and to the high resolution of MIRI at their wavelengths. We further explain the connection of the gas emission with the jet in the sub-section that follows.

\begin{figure}[ht!]
\centering
\includegraphics[width=\textwidth]{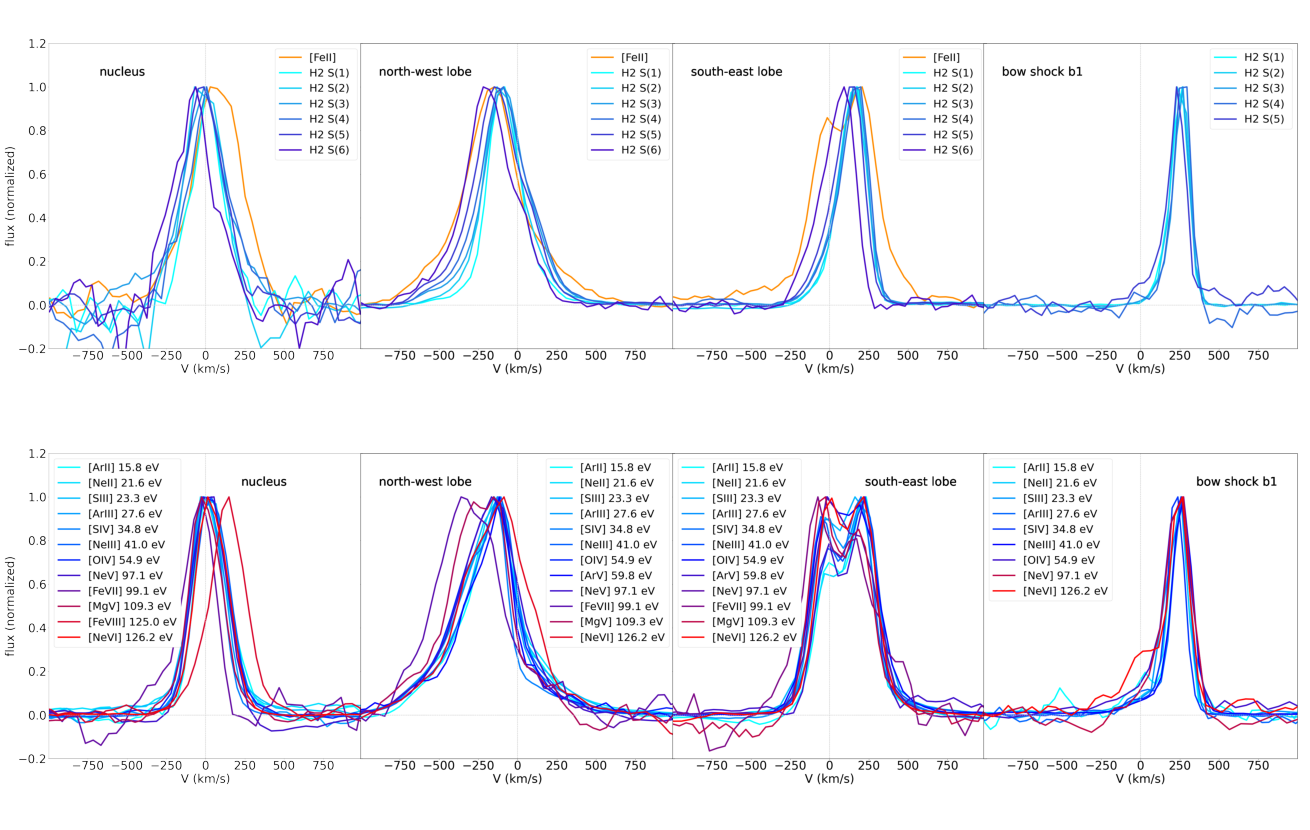}
\caption{Line profiles indicative of  outflows. All spectra are extracted within 1\arcsec\ from the indicated location. {\it Upper row panels:} Normalized spectra of the shocked gas tracers \htwo\ and [\ion{Fe}{2}]. {\it Lower row:}  Normalized spectra of other ionized gas lines, sorted by IP. Only spectral lines with signal-to-noise ratio $>$30 are plotted, so that line wings can be reliably compared.
\label{fig:outflows}
}
\end{figure}

\begin{figure}[ht!]
\centering
\includegraphics[width=\textwidth, trim=0cm 2cm 0cm 4cm, clip]{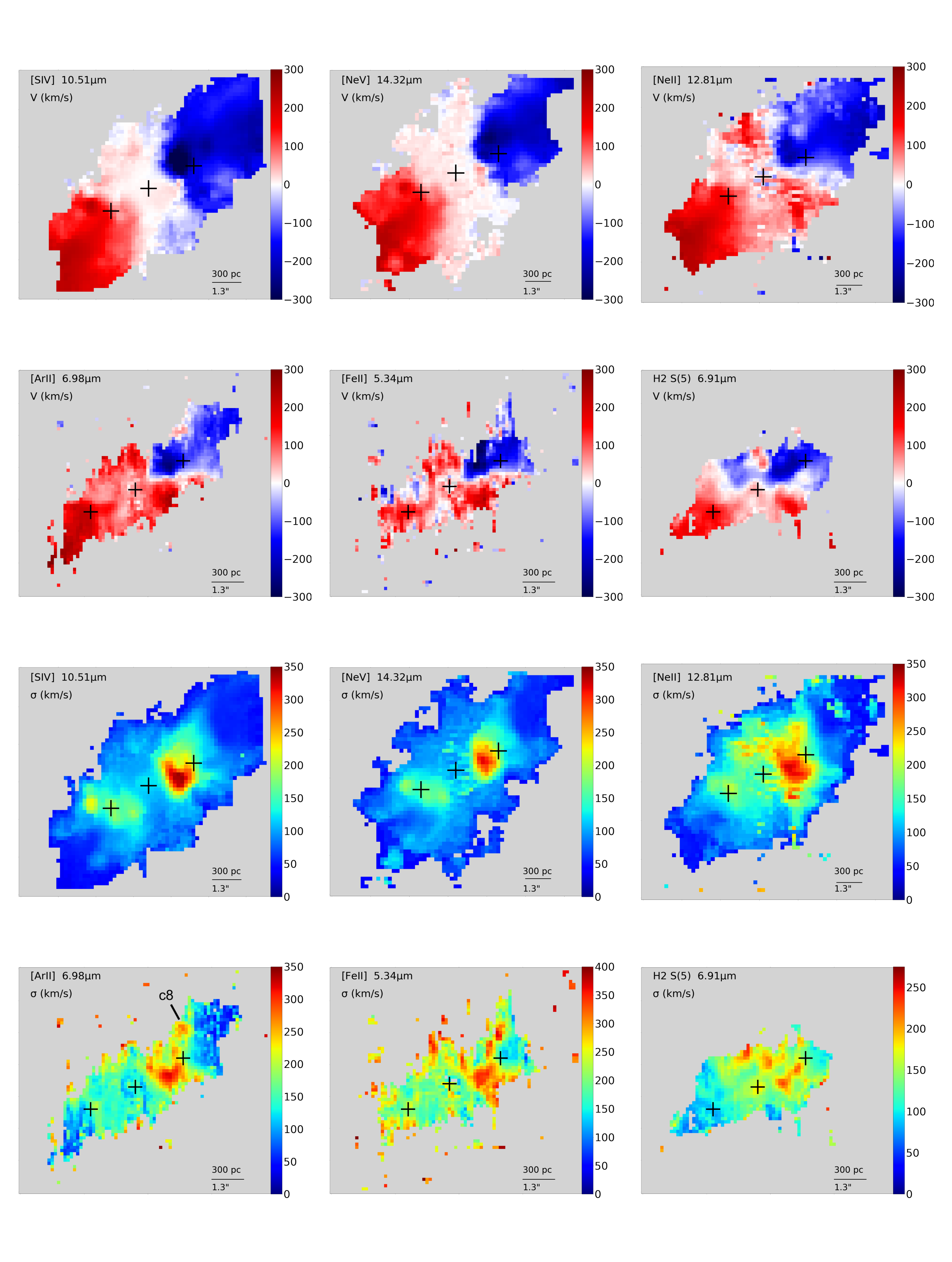}
\caption{ First and second order moment maps (velocity and velocity dispersion) of indicative spectral lines, showing regions with distorted kinematics. Pixels with flux values in the noise were not taken into account for the creation of these maps.
\label{fig:moments}}
\end{figure}

\subsection{Regions influenced by the jet and outflows}
\label{sec:jet-impacted}

Bow-shock-shaped structures are revealed for the first time for this galaxy, already for velocities as low as those of the gaseous disk, i.e., 260\kms\ (as deduced from the \htwo\ S(1) first moment map of Fig.~\ref{fig:disk}; see also \citealt{fonseca23}). A particularly prominent such structure is the \htwo -bright region marked as b1 in Fig.~\ref{fig:all_south}, 1200\,pc south-east of the nucleus. Even though in the low-resolution \htwo\ S(1) data it appears like a blob (Fig.~\ref{fig:disk}), in lines with higher resolution data, i.e. in \htwo\ S(3), S(4), and S(5), its shape is very clearly indicative of a bow shock (b1 in Fig.~\ref{fig:all_south}) that is headed outwards in the direction of the SE branch of the ionized gas EELR (marked as f3 in Fig.~\ref{fig:all_south}). This structure is indeed a bow shock, not a projection effect of spatially unrelated clumps. The bow shock maintains its shape in all of \htwo\ S(3), S(4), and S(5), i.e., in cubes in which there is no detectable emission by any other similar-velocity structure in its vicinity: the disk emission, including some weak spiral arm emission, has already faded by \htwo\ S(3). No associated stellar component can be seen in optical and near infrared data at the location of b1: this is a gaseous structure seen in \htwo\ only. Another bow shock is b2, with a similar orientation according to \htwo\ S(3), S(4), and S(5). In \htwo\ S(7), a third bow-shock-shaped structure is also seen at the SE lobe itself, at $\sim$150 \kms\ (although this structure may be part of the ionized gas bubble b3; see below). The larger-scale structure connecting b1, b2, and b3, as best seen for \htwo\ S(1) in the velocity range 100$<$V$<$400\,\kms , does not follow the spiral patterns described above (marked as sp in Fig.~\ref{fig:all_south}), but the jet axis and the EELR edge instead. These findings suggest that the jet has traveled in the EELR and driven bow shocks in regions b1 and b2. In b1 and all along the f3 filament leading up to b1 (shown e.g., in the \neiii\ image at V$>$300 \kms ; Fig.~\ref{fig:all_south}), a weak outflow is seen in many emission lines (Fig.~\ref{fig:outflows}). In the north-west part of the galaxy, both the edges (f5, f6, f8;  Fig.~\ref{fig:all_north})  and the interior of the ionized gas EELR (f7 and f9) are rather filamentary. One of these structures, f9, is in a nearly straight line with the jet axis, as seen at 17.8 GHz, again suggesting that the jet has propagated in the EELR.

In the 17.8 GHz radio image presented by \citet{morganti07}, bright radio emission exists mainly at the radio core, lobes, and between them. No significant emission is detected near b1. In the other side of the galaxy, contours shown close to the NW radio lobe are elongated in a direction comparable to that of the outer part of the f5 filament, at a PA of $\sim$10$\degree$, as it is seen in  \mgv\ (Fig.~\ref{fig:all_north}) . A similar elongation is also seen in contours of a 8\,GHz image \citep{morganti98}. Therefore, it is possible that some weak radio emission exists, to be revealed with higher sensitivity observations. It is also possible that a jet passed through the ISM, in the past, with a relativistic speed,  and that we now see gas pushed and excited by the slower-moving outflows that it raised. As shown in studies of other nearby galaxies, it is possible to detect the passage of a past or of a weak radio jet by the impact it leaves on the ISM:  \citet{aalto16,aalto20}, \citet{fernandez20}, \citet{pereira-santaella22}, and \citet{papachristou23} demonstrated that outflows can reveal the existence of jets, even if these jets are not seen in radio wavelengths. As we will show below for IC5063, outflows exist at the nucleus, at the radio lobes and in many other regions, including the b1 bow shock. 

Spectra of the outflows seen in these three regions, which are frequently studied in the literature, can also be found in Fig.~\ref{fig:outflows}. Overall, very wide line profile wings are seen close to the radio lobes, and in particular near the NW lobe, where the outflowing ionized gas reaches projected velocities of 1000\kms\ for many ionized gas lines. The widths of the lines with high IP (around or above 100 eV) are wider, showing more prominent outflow components than [\ion{Ar}{2}] with 15.8 eV. The molecular gas attains lower velocities than the ionized gas there, up to 750\kms . Still, the higher the \htwo\ rotational number, the more blueshifted the emission is.
At the SE radio lobe, a similar trend is observed for the molecular gas, and a second blueshifted component associated with an outflow becomes brighter with increasing IP of the ionized gas species. 
These results indicate stratification of the gas in outflows, meaning that the closer the gas is to the power source, the higher its excitation/ionization and its velocity are (e.g., \citealt{armus23}). The stratification of gas in outflows was discovered already in {\it Spitzer} Space Telescope data of local AGN with fast  ionized gas outflows (\citealt{dasyra11a}; see, for example, IRAS05189-2524 and IRAS15001+1433). It is now found to also be possible in regions within galaxies with jet-driven outflows.

An examination of the cube slices at different velocity ranges indicates that there are, in fact, several starting points of outflows. In the overall MIRI FOV, twice as many spatially discrete outflows as those already known from the VLT SINFONI data are now revealed \citep{dasyra15}. In the western part of the galaxy, the fastest outflow with positive velocities (of $\sim$ 1000\kms) is seen in the clump c2 (Fig.~\ref{fig:all_south}), which is about midway between the nucleus and the NW radio lobe, and which lies at the center of an overpressurized region in the MUSE data \citep{dasyra22}. The second highest velocities are found in the clump c1, which is 100($\pm$10)\,pc away from the bifurcation point at the EELR base. 
At negative velocities, of about -1000\kms , yet another clump (marked as c7 in Fig.~\ref{fig:all_north}) is seen in \siv\ 90\,pc north east of the same bifurcation point, and south of a region with high ratio of outflowing to regularly rotating gas seen in SINFONI \feii\ 1.644 $\mu m$ data \citep{dasyra15}.  All these regions could either be discrete starting points of outflows or fragmented clouds in the same outflow. Beyond the bifurcation point, a filamentary outflow is detected along the NW EELR edge (see e.g., \siv\ at -850$<$V$<$-300 \kms\ and \mgv\ at -600$<$V$<$-300 \kms\ in Fig.~\ref{fig:all_north}). Similar findings can be reported for the eastern part of the galaxy. The outflow along the f3 filament in the EELR SE edge extends out to 800\,pc from the SE bifurcation point (see \neiii\ at 300$<$V$<$500 \kms\ in Fig.~\ref{fig:all_south}). An individual clump can be identified in it (c4). In two more high velocity clumps, one close to the f4 edge of the SE EELR (clump c5), and another close to the jet axis (clump c6), the velocities reach 900\kms\ for \siv . Another outflowing clump closer to the nucleus is c3, seen up to 500\kms\ for the same line.

\begin{figure}[ht!]
\centering
\includegraphics[width=0.63\textwidth]{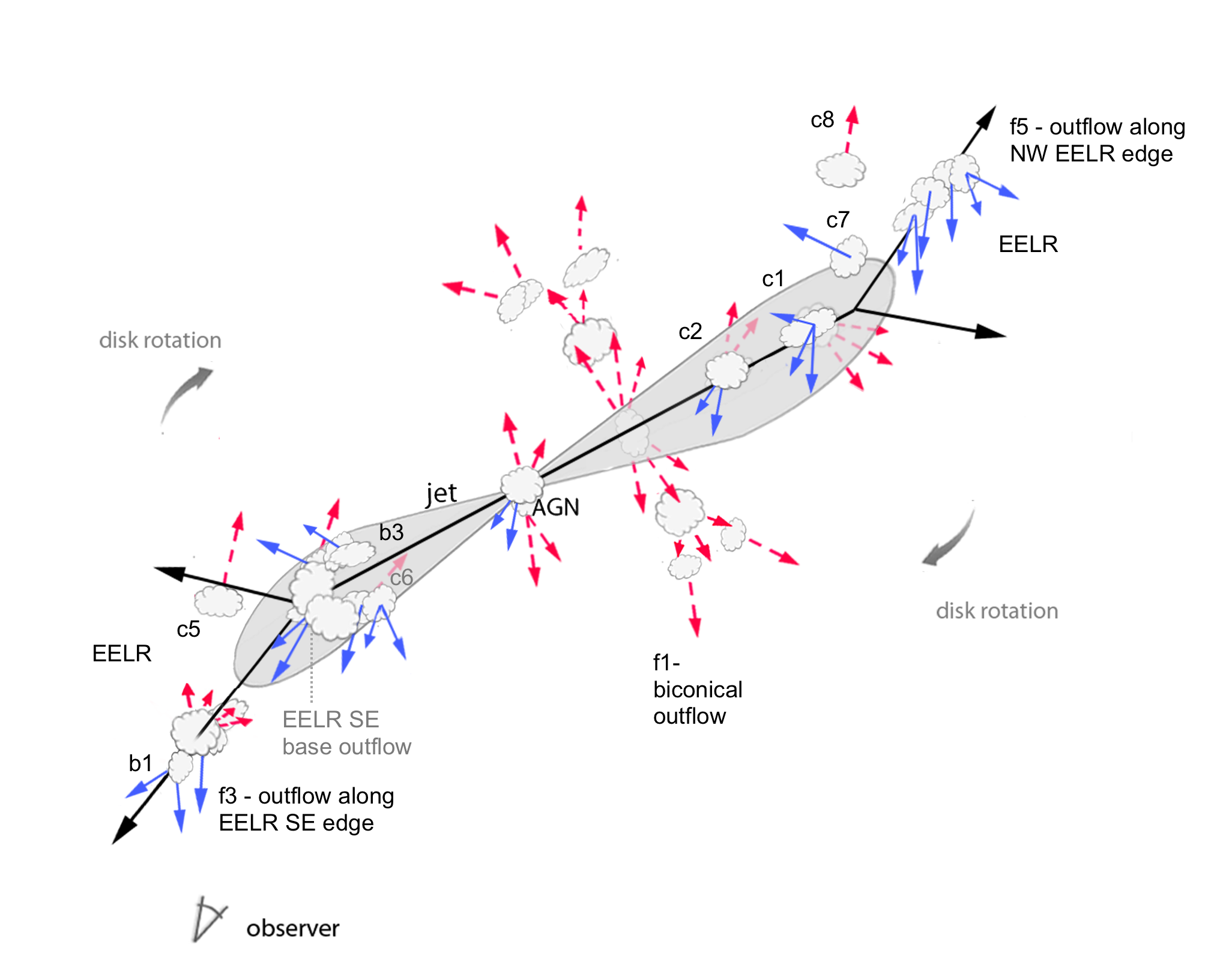}
\caption{Summary, shown as a schematic representation, of all identified regions with outflows. The nomenclature follows that in Figs.~\ref{fig:all_south} and~\ref{fig:all_north}. The disk plane is  in grey. The distances and number of clouds per region are indicative. Red and blue arrows indicate outflows receding from the observer or approaching the observer, respectively.
\label{fig:cartoon}}
\end{figure}

\begin{figure}[ht!]
\centering
\includegraphics[width=0.8\textwidth]{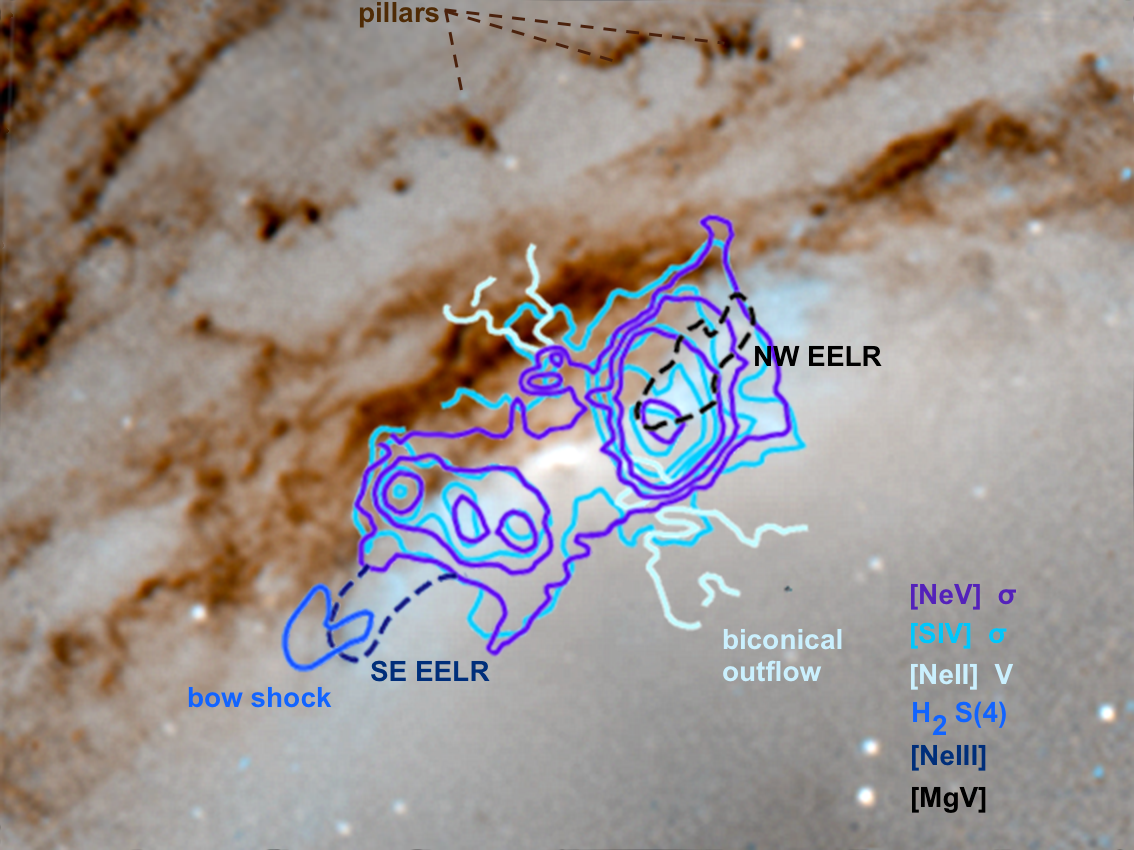}
\caption{ The area in the nucleus of IC5063 with kinematically distorted gas, as seen by the JWST. Highly turbulent regions are indicated with the aid of the velocity dispersion contours of two bright lines of Fig.~\ref{fig:moments}: \siv\ and \nev . Contours are shown at 100, 140 170, 220, 310\kms\ for \siv\ and 110, 130, 160, 290\kms\ for \nev . The biconical outflow is shown with the aid of the velocity contours of \neii , for V=20\kms . The outflows along the NW and SE EELR edges are shown with contours of the \mgv\ -600$<$V$<$-300 \kms\ flux image and of the \neiii\ 300$<$V$<$500 \kms\ flux image, with respective contour levels of 3.5$\times$10$^{-21}$ and 9$\times$10$^{-21}$ W m$^{-2}$. For \mgv , the contours are selectively plotted for outflows at and beyond the NW radio lobe. The \neiii\ contours are plotted further away than the SE radio lobe, in regions not already comprised in the high-$\sigma$ areas, because the outflow and the disk cannot be spatially disentangled close to the EELR base. The b1 bow shock contours are created from the \htwo\ S(4) 220$<$V$<$360\kms\ image,  for a flux of 1.8$\times$10$^{-21}$ W m$^{-2}$.  
 All contours are plotted over the optical image of the galaxy that is shown in Fig.~\ref{fig:FOV}, from NASA / ESA / CSA / J. Schmidt.
\label{fig:outflows_sigma_summary}}
\end{figure}  

A large-scale region that contains gas that is outflowing, moving away from us, is found on the western side of the galaxy. It begins from its own discrete starting point about mid-way between c2 and the nucleus. It extends perpendicularly to the jet propagation axis, both to the north-east and to the south-west of it for at least 500\,pc in each direction. It is marked as f1 in Fig.~\ref{fig:all_south}, and it is seen in velocities in the 0$<$V$<$400\,\kms\ range, with a peak at 220$\pm$30 \kms. Its emission is redshifted, while that of the disk is blueshifted in the same region. It is most prominent in shock-tracing lines, i.e., in all \htwo\ lines from S(3) to S(7) and in \feii\ 5.34 $\mu m$ (Fig.~\ref{fig:all_south}). It has several discrete clumps identified in it. It can also be seen in a few ionized gas lines of low IP, such as \arii . This is the structure that we called biconical outflow in the SINFONI data, in which we first identified it \citep{dasyra15} because the fraction of intermediate-velocity to low-velocity gas in it opens up as a cone with increasing distance from the jet axis. In the SINFONI data, the biconical outflow was seen again in shocked gas tracers, and more specifically, in the ro-vibrational \htwo\ (1-0) S(1) and S(3) lines and \feii\ 1.644 $\mu m$. In the MIRI data, the biconical shape makes this outflow easily distinguishable in the first moment maps of several lines, including those of \htwo\ S(5), \feii , \arii , and \neii\ (Fig.~\ref{fig:moments}). High velocity dispersion is associated with this outflow too, with the greater turbulence detected in the southern part of the galaxy (Fig.~\ref{fig:moments}). In optical MUSE data, \citet{venturi21} had found large-scale turbulence of the ionized gas, probed by an effective width of the \oiii\ line, that extends up to 4 kpc away from the jet axis in both directions. These might be different scale components of the same outflow, or similar orientation outflows with neighboring starting points, generated as the jet collided with neighboring clouds and scattered gas perpendicular to its path. In the eastern part of the galaxy, a similar structure, but much weaker in emission is found. It is marked as f2 in Fig.~\ref{fig:all_south} and it is near regions with negative velocity components south-east of the nucleus in the \arii\ velocity map.

The moment maps also indicate the existence of another clump moving at opposite velocities compared to those of the disk. It is a cloud 300\,pc north of the NW EELR base, seen in both the first and the second moment maps of \arii , \neii, and \feii . It is also seen in the dispersion map of the 1.644 $\mu m$ \feii\ line observed with SINFONI. We call it c8 in Fig.~\ref{fig:moments} (see the \arii\ $\sigma$ panel). 

Finally, one more newly discovered and prominent structure, identified for the first time in the JWST data, is a very bright and clearly delinated bubble (b3 in Fig.~\ref{fig:all_north}). It is located between the SE lobe and the nucleus, and it can be seen for many ionized gas lines of different IPs, even for HI Pfund $\alpha$. It is indicatively shown in Fig.~\ref{fig:all_north} for the bright \feii , \ariii , and \mgv\ lines and for negative velocities down to -300\kms , displaying the bulk of its emission. For negative velocities, its material is approaching us while it is moving against the regular disk rotation. Still, the bubble also contains some gas moving away from us, seen mostly in the 0$<$V$<$100\,\kms\ range. Overall, it is a structure with irregular kinematics moving at low projected velocities.  This structure emits also in \feii\ 1.644 $\mu m$ in SINFONI data, where it appears more like a blob due to the lower resolution of those data. This bubble is likely tracing a jet-driven or a supernova-driven outflow centered mid-way between the nucleus and the SE radio lobe (with a few bright clumps in it). In the latter scenario, it would make IC5063 a great candidate for studying jet-induced star formation, as this supernova could be related to jet-induced star formation.

In total, we see more than ten outflowing clouds and structures in our data. The exact number depends on whether some clumps can be considered part of a common outflow (that started from the same point and that is headed in the same directions) or not, as the medium that we see is very clumpy. A series of jet-cloud interactions and reformation of clouds in outflows (\citealt{zubovas14}) could be contributing to the clumpiness of the medium. We note that it is possible that even more outflows exist in our data, but that the cruciform artifact (see Section~\ref{Appendix:cruciform}) prevents us from identifying them with certainty. Such an example is the structure u1 (Fig.~\ref{fig:all_south}). Similarly, some low S/N regions seen around the nuclear PSF in the -1150$<$V$<$-850\kms\ \siv\ image could either be real or due to PSF residuals.

We summarize all unambiguously identified regions with outflows in our data in Fig.~\ref{fig:cartoon}, using a schematic representation. In this representation, the jet travels in the ISM and drives both receding and approaching outflows in regions of direct collision with dense clouds, where gas is dispersed in all directions. Outflows characterized by redshifted emission only originate from clouds that are located behind the radio jet with respect to the observer and that are pushed further away by the jet. Inversely, outflows characterized by blueshifted emission only originate from clouds that are located in front of the radio jet with respect to the observer and that are pushed closer to the observer by the jet. This explanation holds irrespective of which galactic hemisphere each outflow is found in. 

A plot visualizing all regions with high gas turbulence can be found in Fig.~\ref{fig:outflows_sigma_summary}. It shows how wide-spread the turbulence in the center of IC5063 is. In total, the gas kinematics are disturbed a region of 2.2$\times$1.6 kpc$^2$. Moreover, turbulent medium appears to be moving in projection towards a dusty structure seen in the optical image of Figs.~\ref{fig:FOV} and \ref{fig:outflows_sigma_summary} that runs nearly parallel to the jet axis, $\sim$1200\,pc north of it. This dusty structure is seen as a wavy interface and pillar-shaped clouds, which could be associated with radiation pressure or ram pressure (and associated instabilities, such as the Vishniak instability). Winds travelling along the biconical outfow, or along the axis indicated by the \nev\ and \siv\ velocity dispersion contour elongation (i.e., at a $-$15$\degree$ position angle from the NW radio lobe), could have acted on this structure, giving it its current shape. It is thus possible that these clouds have been affected by a past episode of jet or AGN radiation feedback.

\begin{figure}[ht!]
\centering
\includegraphics[width=\textwidth]{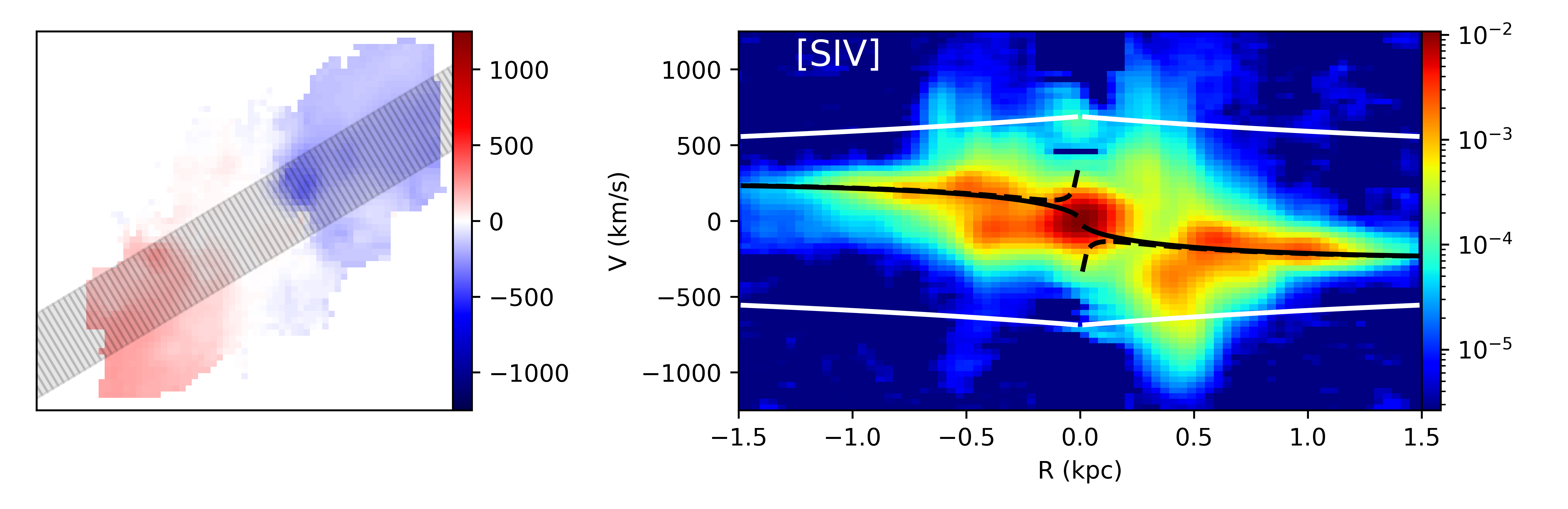}\\
\caption{ {\it Left:} Pseudo-slit position angle and width used for the position-velocity diagram creation, plotted over the \siv\ moment 1 map. The position angle is identical to the major axis of rotation, as derived from the \htwo\ S(1) rotation field. {\it Right:} Position velocity diagram along the major axis of rotation. Pixel values below noise have been masked in this diagram. The black solid line shows the rotational velocity in the Hernquist bulge that best characterizes this galaxy (see the text for more details). The black dotted line shows for completion the same velocity because of the black hole potential. The positive and negative escape velocities from the Hernquist bulge are also plotted as a function of radius, using white solid lines. An inclination correction is applied to all theoretical velocities, for consistency with the observations.
\label{fig:pvd}}
\end{figure}

Another diagram showing how ubiquitous outflows are in our data is the position-velocity diagram along the major axis of rotation, constructed for the bright \siv\ line (Fig.~\ref{fig:pvd}). Theoretical curves of the rotational and escape velocities as a function of radius are also overplotted. The gravitational potential associated with the matter distribution in IC5063, at the radii under examination, is assumed to be well approximated by a Hernquist potential \citep{mukherjee18}, for which analytic formulas describe the radial profiles of the rotational velocity and the escape velocity  \citep{hernquist90}. For the mass and the scale length of the stellar distribution needed for this calculation, we also adopted the parameter values of \citet{mukherjee18}. These relied on observations of the central stellar velocity dispersion ($160\,\mathrm{km\,s^{-1}}$) and effective radius ($21.5\,\mathrm{\arcsec}$) from \citet{kulkarni98} and \citet{morganti15}, and indicated a stellar mass of $1.67\times10^{11}\,\mathrm{M_\odot}$ and a scale length of $2.8\,\mathrm{kpc}$ for the stellar bulge. Correcting for the inclination of the galaxy, with an inclination angle of $i=74^\circ$ \citep{morganti15}, we obtained the rotational velocity and escape velocity curves in the position-velocity diagram (Fig.~\ref{fig:pvd}). The position angle along which the \siv\ flux was collapsed was identical to that of the major axis of rotation of \htwo\ S(1), 108$^\circ$ (Fig.~\ref{fig:disk}). Unlike the molecular gas traced by \htwo\ S(1), which never reaches projected velocities greater than the escape velocity, the outflowing ionized gas probed by \siv\ indeed travels faster than the escape velocity nearly all along the inner 600 parsec of the galaxy. Only in the $\sim$100 central parsecs this claim cannot be reliably made due to PSF residuals and the to contribution of the black hole of 2.8$\times$10$^8$\msun\ (\citealt{nicastro03}) to the gravitational potential. In the south-east of the galaxy, outflowing gas is continuously distributed from 600 to 1500\,pc from the nucleus, appearing all along the EELR edge. It can be seen in the position-velocity diagram and in Fig.~\ref{fig:outflows_sigma_summary} with the aid of the \neiii\ contours leading up to the b1 bow shock. Weak line wings associated with this nearly-kiloparsec-long outflow can be seen in Fig.~\ref{fig:outflows}, 
near b1. There, the projected gas velocities  do not reach the corresponding escape velocities. At $\sim$600\,pc, the other end of this outflow,  a blue wing in \oiii\ 5007\ang\ has long been known to exist \citep{wagner89}. At distances below 600\,pc, the filament f10 seen for \htwo\ S(5) (Fig. \ref{fig:all_north}), which is nearly parallel to the jet axis, likely contains outflowing gas.

\subsection{Density of the ionized gas}
\label{sec:density}

\begin{figure}[ht!]
\centering
\includegraphics[width=0.45\textwidth]{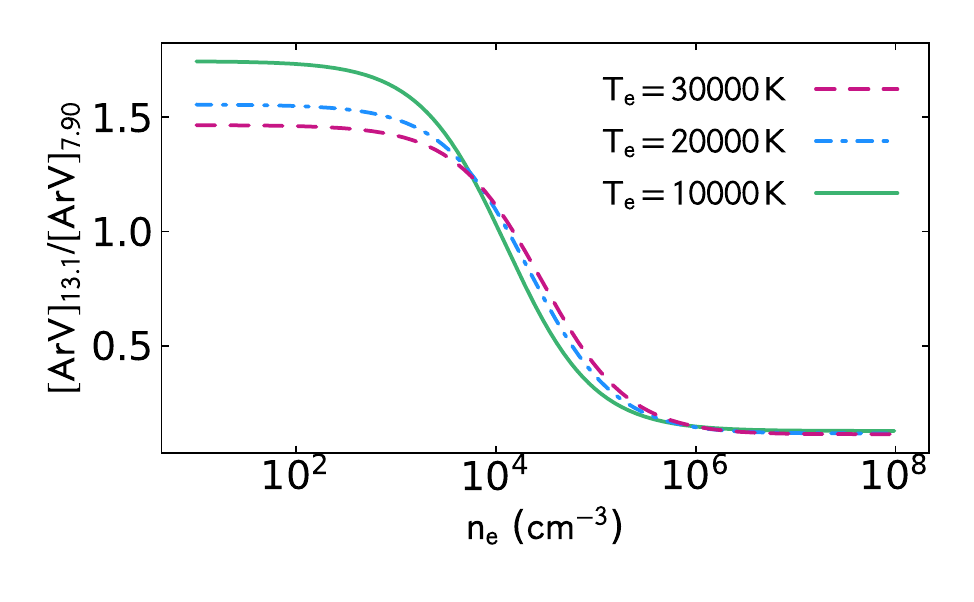}
\includegraphics[width=0.33\textwidth]{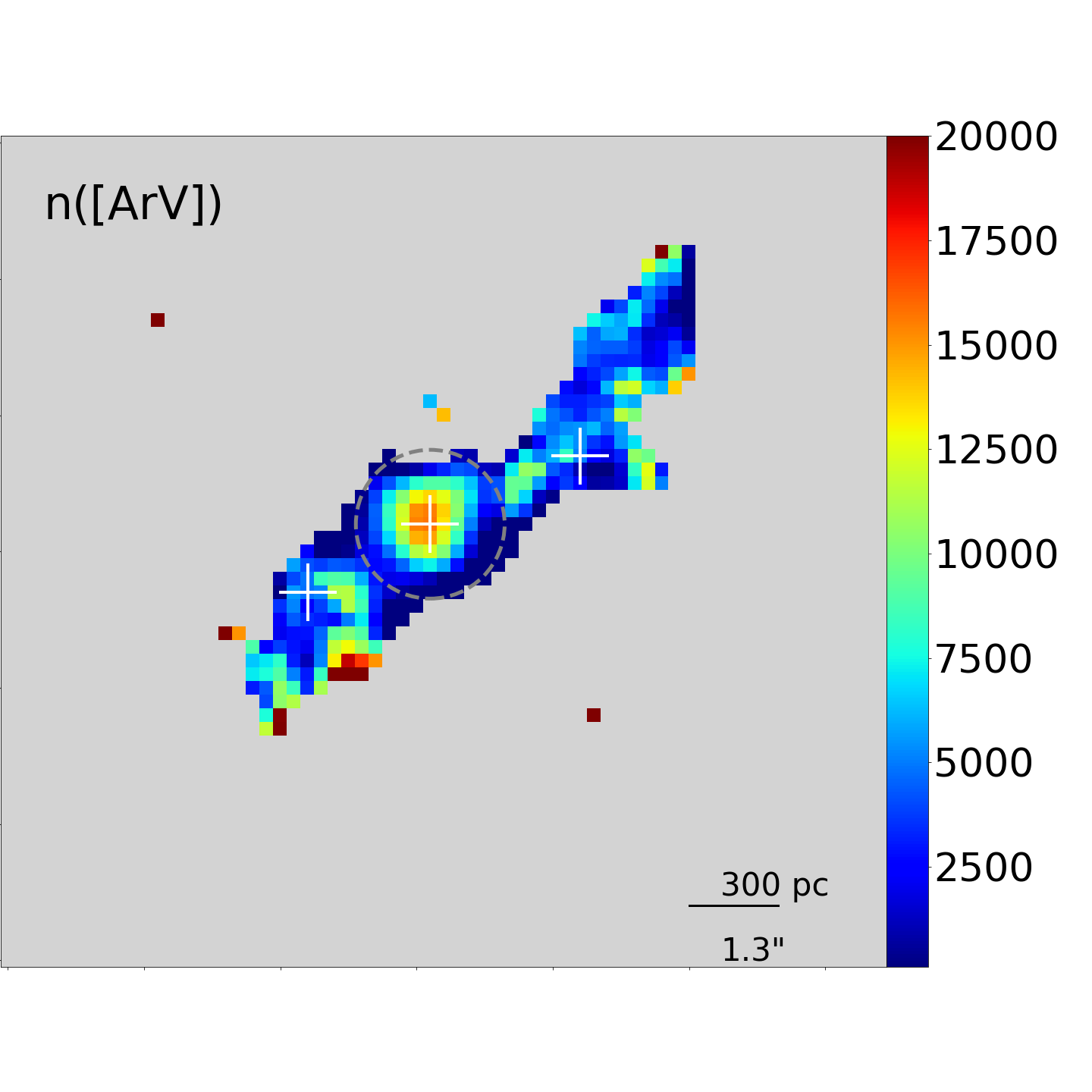}\\

\caption{ {\it Left:} The \arv\ line ratio as a function of electron density in LTE conditions for different temperatures. {\it Right:} \arv -based electron density map. Crosses are as in the previous figures. The dashed circle indicates the region where the density values are uncertain due to PSF residuals. Densities are in units of cm$^{-3}$.
\label{fig:density_probes}}
\end{figure}

The density of the ionized gas can be determined in the MIRI wavelength range with two sets of lines of the same species and ionization potential: the \arv\ 13.1$\mu m$/7.9$\mu m$ flux ratio and the \nev\ 24.3$\mu m$/14.3 $\mu m$ flux ratio. We use the former ratio, which has sufficient spatial resolution for the creation of a map. The density range that \arv\ 13.1$\mu m$/7.9$\mu m$ can measure is shown in Fig.~\ref{fig:density_probes}. Its values are computed with the code Pyneb \citep{luridiana15} for different temperatures and for line intensities that are reflecting electronic levels that are populated based on the Boltzmann distribution, i.e., for gas under LTE. The derived density map (Fig.~\ref{fig:density_probes}) was computed for an indicative temperature of 10000K. This temperature was deemed appropriate given that flux ratios $>$ 1.5 were found in several pixels of our data, and also given that previous temperature calculations from [\ion{N}{2}] lines in MUSE data were in the range $\sim$7000-12000K \citep{dasyra22}. The density map shows discrete regions of overdense ionized gas, with densities somewhat higher than 10000 atoms per cubic centimeter in locations where outflow starting points are seen, including  c2. For comparison, for the vicinity of c2, \citet{holden23} report densities of 6000 cm$^{-3}$ based on [\ion{Ar}{4}] in the optical. Similarly, in the NW EELR (regions f6, f7), in a region with outflows in our data and with particular polarization in HST FOC data (potentially due to the outflows; \citealt{barnouin23}), the density seems elevated. Overall, it is noteworthy that the \arv\ diagnostic ratio, which is applicable at rather high density values ($>$1000 cm$^{-3}$; Fig.~\ref{fig:density_probes}) and ionization potential (59.8 eV), works well for so many regions of the nucleus of IC5063. This result could plausibly be indicative of compression of the ionized medium.

\begin{table}[h!]
\begin{flushleft}
{\caption{Line properties}}
\begin{tabular}{l l l l} 
\hline\hline 
Line name  & $\lambda$ & f($^a$) & $\sigma$
 \\ 
   (-) &	($\mu$m) & (10$^{-17}$W/m$^2$) & (\kms ) \\  
\hline 
H2 S(8) &  5.053 & 1.28  ($\pm$ 0.32) & 130  ($\pm$ 12) \\ 
\lbrack FeII\rbrack  &  5.340 & 5.64  ($\pm$ 0.62) & 176  ($\pm$ 13) \\ 
\lbrack FeVIII\rbrack  &  5.447 & 3.17  ($\pm$ 0.11) & 84  ($\pm$ 9) \\ 
\lbrack MgVII\rbrack  &  5.503 & 3.52  ($\pm$ 0.56) & 75  ($\pm$ 8)\\ 
H2 S(7) &  5.511 & 3.84  ($\pm$ 0.52) & 150  ($\pm$ 30) \\ 
\lbrack MgV\rbrack  &  5.610 & 5.08  ($\pm$ 0.48) & 90  ($\pm$ 13) \\ 
\lbrack KIV\rbrack  &  5.982 & 0.52  ($\pm$ 0.10) & 135  ($\pm$ 15) \\ 
H2 S(6) &  6.110 & 1.68  ($\pm$ 0.29) & 113  ($\pm$ 20) \\ 
\lbrack CaVII\rbrack  &  6.154 & 0.46  ($\pm$ 0.11) & 71  ($\pm$ 11) \\ 
PAH 6.2 &  6.220 & 11.87  ($\pm$ 0.84) & ... \\ 
\lbrack SiVII\rbrack  &  6.492 & 1.36  ($\pm$ 0.35) & 189  ($\pm$ 31)  \\ 
\lbrack NiII\rbrack  &  6.636 & 1.33  ($\pm$ 0.29) & 243 ($\pm$ 39) \\ 
H2 S(5) &  6.910 & 9.26  ($\pm$ 0.87) & 129  ($\pm$ 13) \\ 
\lbrack ArII\rbrack  &  6.985 & 9.44  ($\pm$ 1.80) & 119  ($\pm$ 19) \\ 
\lbrack NaIII\rbrack  &  7.318 & 1.26  ($\pm$ 0.18) & 138  ($\pm$ 13) \\ 
HI Pfund $\alpha$ &  7.460 & 1.93  ($\pm$ 0.30) & 144  ($\pm$ 23) \\ 
\lbrack NiI\rbrack+PAH 7.7($^b$)  &  $\sim$7.5 & 1.10  ($\pm$ 0.30) & ... \\ 
\lbrack NeVI\rbrack  &  7.652 & 33.83  ($\pm$ 1.27) & 80  ($\pm$ 8) \\ 
\lbrack FeVII\rbrack  &  7.815 & 1.44  ($\pm$ 0.24) & 96  ($\pm$ 9) \\ 
\lbrack ArV\rbrack  &  7.902 & 2.60  ($\pm$ 0.44) & 188  ($\pm$ 15) \\ 
H2 S(4) &  8.025 & 4.60  ($\pm$ 0.66) & 128  ($\pm$ 5) \\ 
\lbrack ArIII\rbrack  &  8.991 & 11.83  ($\pm$ 0.99) & 112  ($\pm$ 13) \\ 
\lbrack NaIV\rbrack  &  9.041 & 0.63  ($\pm$ 0.09) & 250 ($\pm$ 60) \\ 
\lbrack FeVII\rbrack  &  9.527 & 3.25  ($\pm$ 0.38) & 174  ($\pm$ 35) \\ 
H2 S(3) &  9.665 & 13.34  ($\pm$ 1.40) & 130  ($\pm$ 7) \\ 
\lbrack SIV\rbrack  &  10.511 & 54.13  ($\pm$ 3.70) & 121  ($\pm$ 13)\\
PAH 11.3 &  11.3 & 30.77  ($\pm$ 6.05) & ... \\ 
H2 S(2) &  12.279 & 5.67  ($\pm$ 0.99) & 122  ($\pm$ 16) \\ 
HI Humphrey $\alpha$ &  12.370 & 0.77  ($\pm$ 0.13) & 162  ($\pm$ 21) \\ 
\lbrack NeII\rbrack  &  12.814 & 28.53  ($\pm$ 3.03) & 126  ($\pm$ 12) \\ 
\lbrack ArV\rbrack  &  13.102 & 2.41 ($\pm$ 0.61) & 89  ($\pm$ 5) \\ 
\lbrack NeV\rbrack  &  14.322 & 29.75  ($\pm$ 2.60) & 90  ($\pm$ 10) \\ 
\lbrack NeIII\rbrack  &  15.555 & 79.29  ($\pm$ 5.30) & 125  ($\pm$ 14) \\ 
H2 S(1) &  17.035 & 12.86  ($\pm$ 1.16) & 125  ($\pm$ 17) \\ 
\lbrack FeII\rbrack  &  17.936 & 1.73  ($\pm$ 0.45) & 114  ($\pm$ 24) \\ 
\lbrack SIII\rbrack  &  18.713 & 31.29  ($\pm$ 1.37) & 144  ($\pm$ 14) \\ 
\lbrack NeV\rbrack  &  24.318 & 23.21  ($\pm$ 2.31) & 97  ($\pm$ 8) \\ 
\lbrack OIV\rbrack  &  25.890 & 107.83  ($\pm$ 6.51) & 130  ($\pm$ 10)\\ 
\lbrack FeII\rbrack  &  25.990 & 4.33 ($\pm$ 0.75) & 113  ($\pm$ 33) \\ 
H2 S(0) & 28.219 & $<$3.6($^c$) & ...  \\
\hline 
\end{tabular}
\newline
{\footnotesize
($^a$) Fluxes integrated over the entire FOV. The FOV differs from channel to channel, as shown in Fig.~\ref{fig:FOV}, but this has a small effect for the integrated flux of most lines, as the bulk of the emission lies within the common FOV (Fig.~\ref{fig:all_lines}). The error bar associated with each flux corresponds to the difference between the value derived from the direct integration of the signal on top of the continuum in the spectrum shown in Fig.~\ref{fig:spectrum} and the value derived when spatially integrating the pixel fluxes in collapsed images of continuum-free cubes (Fig.~\ref{fig:all_lines}). This way, each error bar encapsulates uncertainties in the continuum subtraction (and therefore exceeds the noise accumulated over the line profile. ($^b$) These lines cannot be unambiguously distinguished. ($^c$) This line is undetected in all spectral pixels of the cube and in their collapsed image. Given the measured redshift of this galaxy, the \htwo\ S(0) line is anticipated in the very last $\sim$25 spectral pixels of the ch4 long cube. The instrumental throughput was low there, dropping at the time of our observations, which led to a highly uncertain flux calibration. A tentatively measured flux limit of 3.6$\times$10$^{-17}$Wm$^{-2}$ is below the flux provided in \cite{guillard12} from {\it Spitzer} Space Telescope data. Therefore, this line is excluded from all our analyses. }
\label{tab:line_fluxes}
\end{flushleft}
\end{table}

\begin{figure}[ht!]
\centering
\includegraphics[width=0.33\textwidth]{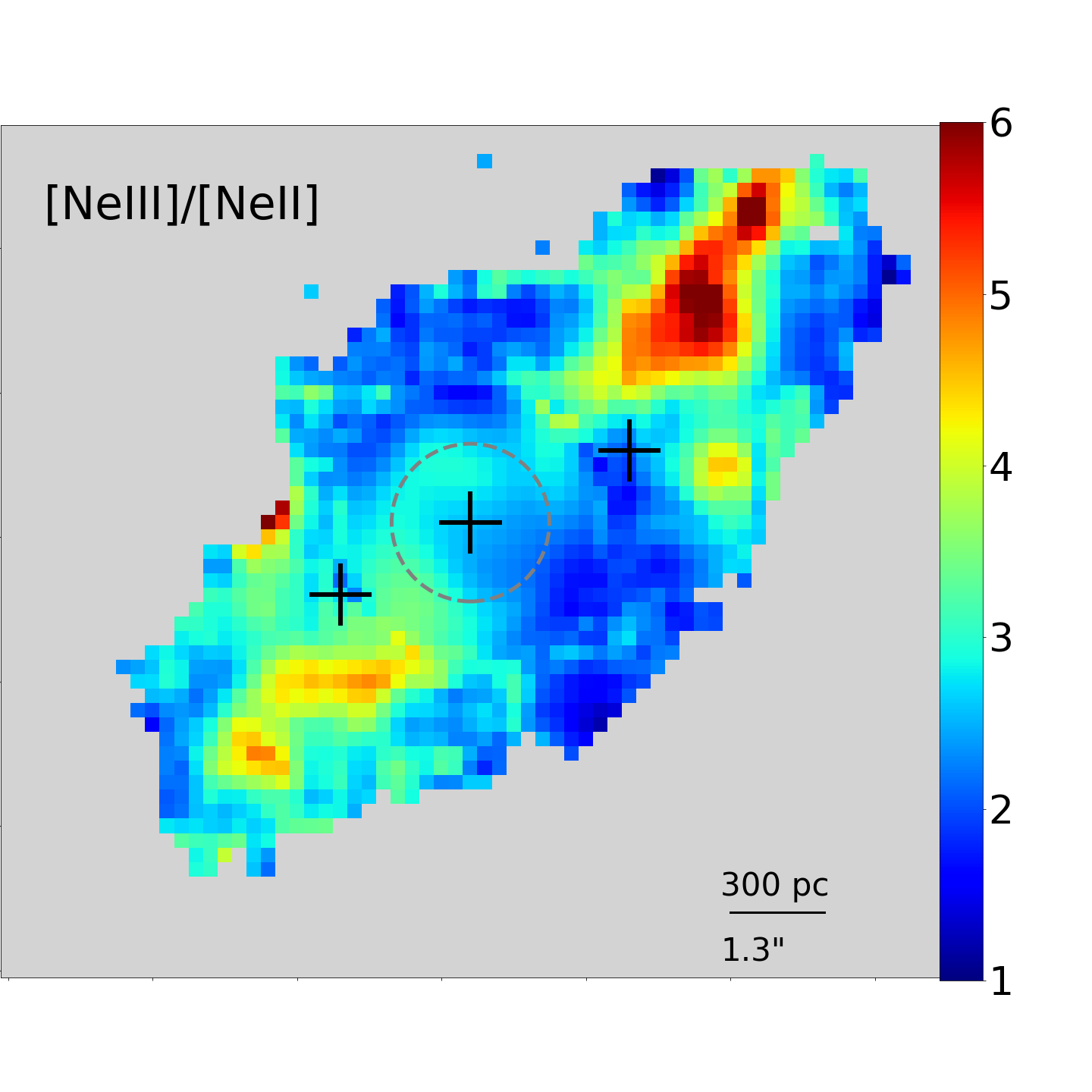}\\
\caption{ \neiii /\neii\ ratio, integrated over all velocities, as an excitation indicator. The circle and crosses are as in Fig.~\ref{fig:density_probes}.
\label{fig:excitation_neon}}
\end{figure}

\section{Discussion: The excitation mechanism(s) of the gas} 
\label{sec:excitation}

\subsection{Ionized gas}
\label{disc:ionized}

In the previous Section, we demonstrated that the jet plays a major role in the observed gas properties in the center of IC5063. A series of outflows along and perpendicular to the known jet axis from radio wavelengths cause great turmoil in the gas kinematics. Bow shocks out to $\sim$1200\,pc and filamentary outflows all along the ionized gas EELR SE branch, as well as outflows out to $\sim$800\,pc along the ionized gas EELR NW branch, suggest that the bifurcating appearance of the EELR in both sides of the galaxy is due to jet scattering at the radio lobes. Our findings indicate that the jet is far more important than the AGN radiation in exciting the gas than previously thought 
\citep{morganti07,mingozzi19, venturi21}. Of course, the AGN radiation may still be contributing to the gas excitation in regions previously cleared out by the jet, as indicated by the existence of AGN radiation and jet inflated cavities in X-ray data \citep{travascio21} and the existence of circumnuclear high-ionization line emission in our MIRI data. Still, the jet contribution to the gas kinematics and excitation is now unambiguously proven even further away than the radio lobes. Already since the presentation of the EELR of IC5063, \citet{colina91} pointed out that the radial profile of the ionized gas emission does not agree with the radial profile of the ionized gas density illuminated by a nuclear point source (i.e., the AGN). Instead, these two radial profiles require the existence of an extended ionization source, i.e., of a radio jet, in order to agree. It is even possible that the jet has traveled at a greater position angle than that of the disk (i.e., at an orientation more inclined towards NW-SE), because the gas emission is brightest in the SE EELR branch east of the nucleus and in the NW EELR branch west of the nucleus. The ionized gas excitation, when examined with the aid of the flux ratio of two bright lines, \neiii\ and \neii\ (shown in Fig.~\ref{fig:excitation_neon}), shows that these two branches also contain the most excited gas (see also a similar result for the [\ion{O}{3}]/H$_{beta}$ ratio in the optical; \citealt{fonseca23}).  Lines of high IP, such as [\ion{Fe}{8}] (with 99.1eV) emit along the same EELR branches in our Webb data, and outflows are detected there too.

\subsection{The molecular gas excitation}
\label{disc:molecular}

\begin{figure}[ht!]
\centering
\includegraphics[width=1.05\textwidth]{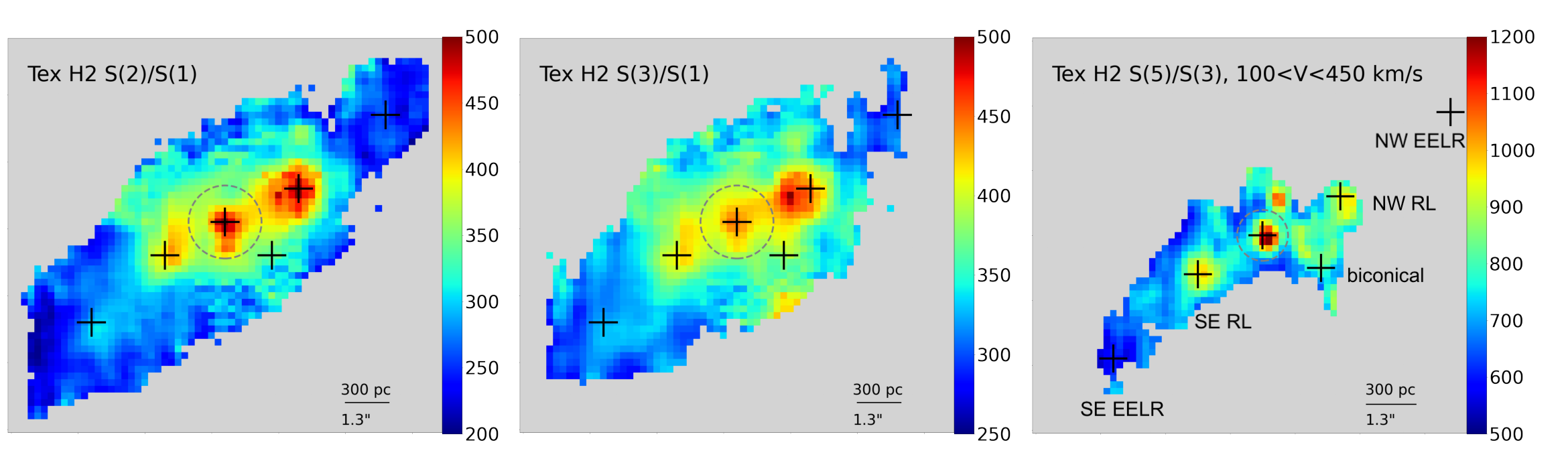}\\
\includegraphics[width=0.5\textwidth]{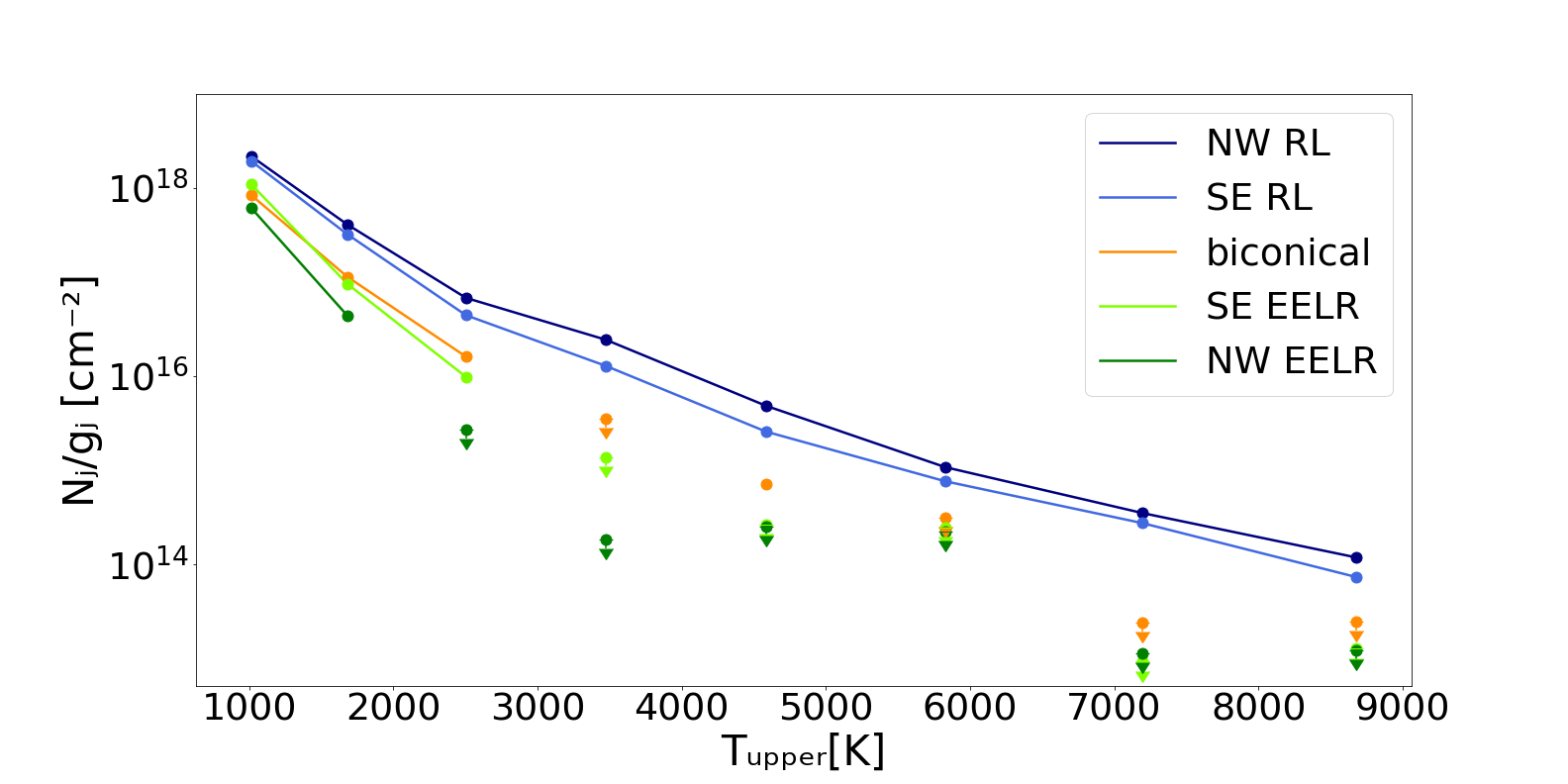}\\
\caption{ {\it Upper panels}: \htwo\ excitation temperature maps. All temperatures are in units of K. Crosses mark the centers of the indicative regions (of 1\arcsec\ diameter) in which line fluxes were measured for the excitation diagram. The central three crosses are identical to those of all previous Figures. The cross at the SE EELR corresponds to the location of the b1 bow shock. The T$_{ex}$ is shown for the integrated line flux over all velocities, unless otherwise noted (i.e., except for the biconical outflow, shown in the third column). {\it Lower panels}: \htwo\ excitation diagram of the above-shown indicative regions.
\label{fig:H2_tex}}
\end{figure}

The warm molecular gas, as seen directly in the \htwo\ rovibrational lines, has been known since {\it Spitzer} observations to show enhanced emission in radio galaxies compared to star forming galaxies thanks to the presence of jet-related turbulence \citep{ogle10,guillard12}. This finding is also excellently demonstrated in our new JWST data of IC5063, in a spatially resolved manner. The excitation temperature of the warm \htwo\ is shown in Fig.~\ref{fig:H2_tex} for indicative line ratios. The computation is made under the assumptions that the line emission originates from gas that is optically thin and in LTE, so that all molecules are probed by photons and so that the Boltzmann distribution applies to the rotational level populations. These are meaningful and widely-applied assumptions for the low rotational levels of the MIR-emitting gas (e.g., \citealt{rigopoulou02,higdon06,roussel07,ogle10,dasyra11b,guillard12,pereira-santaella22}). The flux of two lines is needed for the computation of the excitation temperature T$_{ex}$, which can then be used for the mass computation. T$_{ex}$ can be computed by equating the total number molecules as given by two lines, or equivalently, by calculating the negative of the inverse of the slope between two points in the excitation diagram \citep{rigopoulou02}. The excitation diagram is shown as the natural logarithm of the column density of electrons N$_J$ that descended from an upper to a lower state, divided by the statistical weight g$_J$ of the transition, as a function of the temperature or the energy of the upper state. The number of electrons that made the transition is computed as L/($\alpha$h$\nu$), where h is the Planck constant, $\alpha$ is the Einstein coefficient of the transition, and $\nu$ is the frequency of the emitted line. Multiplied by the pixel size, this number gives the desired column density.

The T$_{ex}$ map for the coolest gas in our data, that emitting in S(1) and S(2), is in many regions quite similar to the same map computed for the ortho lines S(1) and S(3), which are brighter. This is an indication that the low-J lines are thermalized in many regions of our FOV.  The same conclusion can be drawn when examining the full excitation diagram of some indicative regions (Fig.~\ref{fig:H2_tex}; lower panel). This result is meaningful in the context of the gas densities measured so far: when roughly assuming that the \htwo\ gas density is intermediate to that of \arv\ and that of CO (\citealt{dasyra22}), then it is of order 10$^4$ cm$^{-3}$. At such densities, the \htwo\ rotational levels can be rather safely considered thermalized up to at least J=5 \citep{lebourlot99}.

The T$_{ex}$ maps based on \htwo\ S(2) and S(1), S(3) and S(1), as well as S(4) and S(2) show some more intriguing results: that the highest T$_{ex}$ values are seen at the vicinity of the radio lobes and that the temperature is lower by at least $\sim$100\,K in the bow shock regions than in the region where synchrotron emission is detected, within $\sim$500\,pc from the nucleus \citep{morganti98,morganti07}. Interestingly, this inner, synchrotron-emitting region also shows a step in the cold molecular gas heating rate and excitation temperature in such maps derived from CO and HCO$^+$ observations \citep{dasyra22}. The CO density also happens to be elevated in that region, just like the ionized gas density in the optical MUSE data  \citep{dasyra22}. All these findings imply that there is an additional excitation mechanism of the molecular gas that is linked to the jet.

Two jet-related mechanisms are highly likely candidates: either a superposition of shocks or  cosmic rays (see also \citealt{ferland08,nesvadba10,padovani22,leftley24}). Indeed, either mechanical heating or cosmic ray heating or a combination of the two was deemed possible in our CO modeling with the 3D-PDR code \citep{dasyra22}, which self-consistently calculates the formation and destruction rate of molecules \citep{bisbas15,bisbas17}. For the CO, an increase of the cosmic ray ionization rate from about 10$^{-16}$\,s$^{-1}$ to 10$^{-14}$\,s$^{-1}$ was found plausible to reproduce the millimeter data. In the optical, a significant shock contribution was found necessary by \citet{fonseca23} to explain some high \fevii /H$_{\beta}$ and [\ion{Fe}{10}]/H$_{\beta}$ seen in MUSE data.

Both mechanisms are likely to be at play, but shocks seem to be of major importance according to the new \htwo\ data for two reasons. The first reason has to do with the molecular hydrogen cooling rate. If we add the fluxes of the \htwo\ lines in the MIRI data that are primarily responsible for the radiative cooling, i.e., including S(1), S(2), and S(3) \citep[e.g.,][]{ogle10,nesvadba10}, then we get a total luminosity of 9.2$\times$10$^{40}$ erg/s. By further adding the flux of the S(0) line from the {\it Spitzer} IRS data, 4.6$\times$10$^{-17}$ W m$^{-2}$  \citep{guillard12}, then the luminosity reaches 1.0$\times$10$^{41}$ erg/s. Dividing this luminosity with a total mass of 1.0$\times$10$^7$\msun\ (see subsection~\ref{disc:molecularmass}), we get a cooling rate of 1.7$\times$10$^{-23}$ erg/s per molecule. Assuming, per \citet{nesvadba10}, $\sim$12eV per ionization per molecule by cosmic rays, this would require a cosmic ray ionization rate $\zeta_{CR}$ of order 10$^{-12}$ s$^{-1}$ to counterbalance the cooling. Even though the actual $\zeta_{CR}$ could be up to an order of magnitude higher if a more massive reservoir could be detected from similar S(0) observations, (see again subsection~\ref{disc:molecularmass}), this number would still remain high compared to the Galaxy, to local AGN, and to the appropriate $\zeta_{CR}$ values found from the CO modeling \citep{dasyra22}.

\begin{figure}[ht!]
\centering
\includegraphics[width=0.35\textwidth]{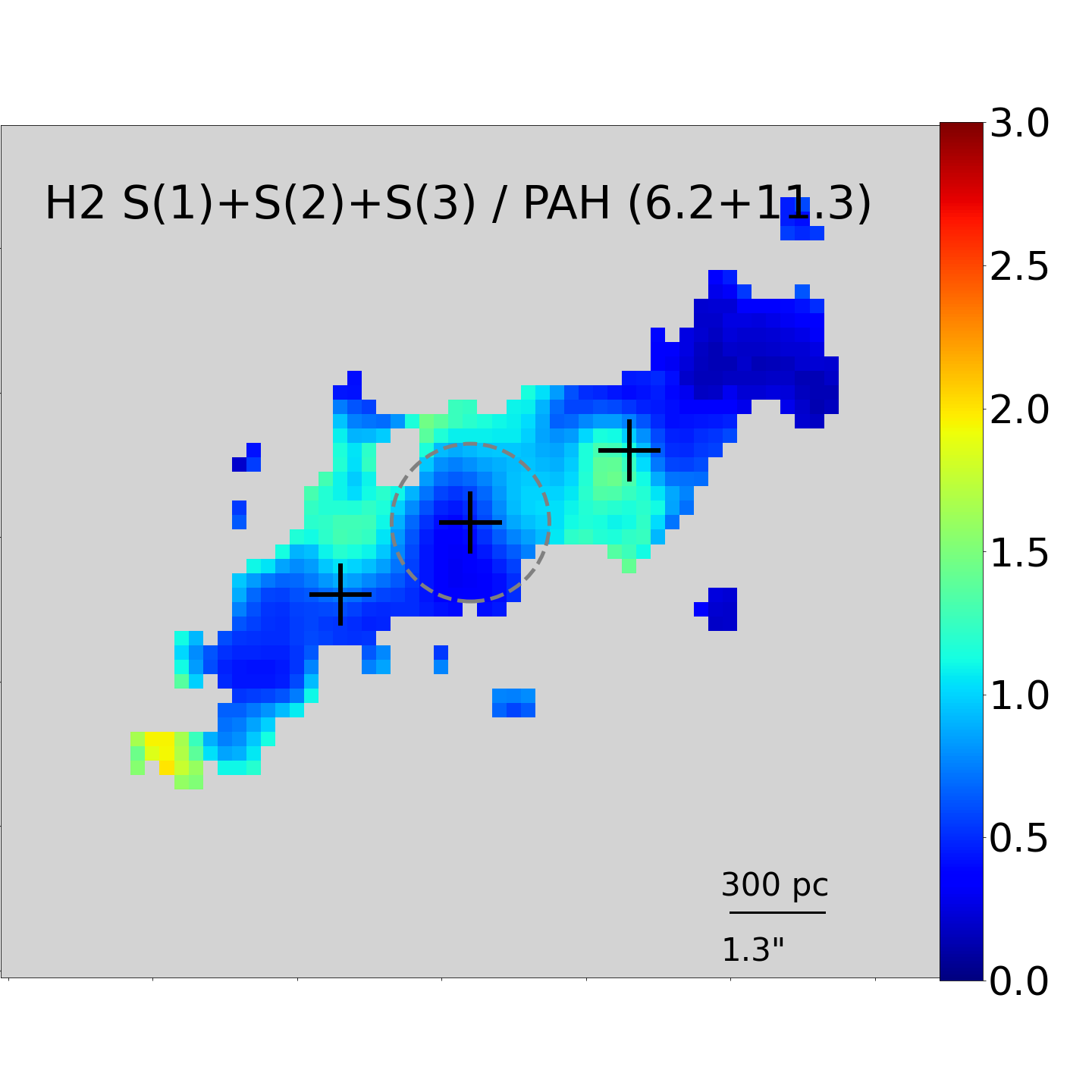}
\includegraphics[width=0.35\textwidth]{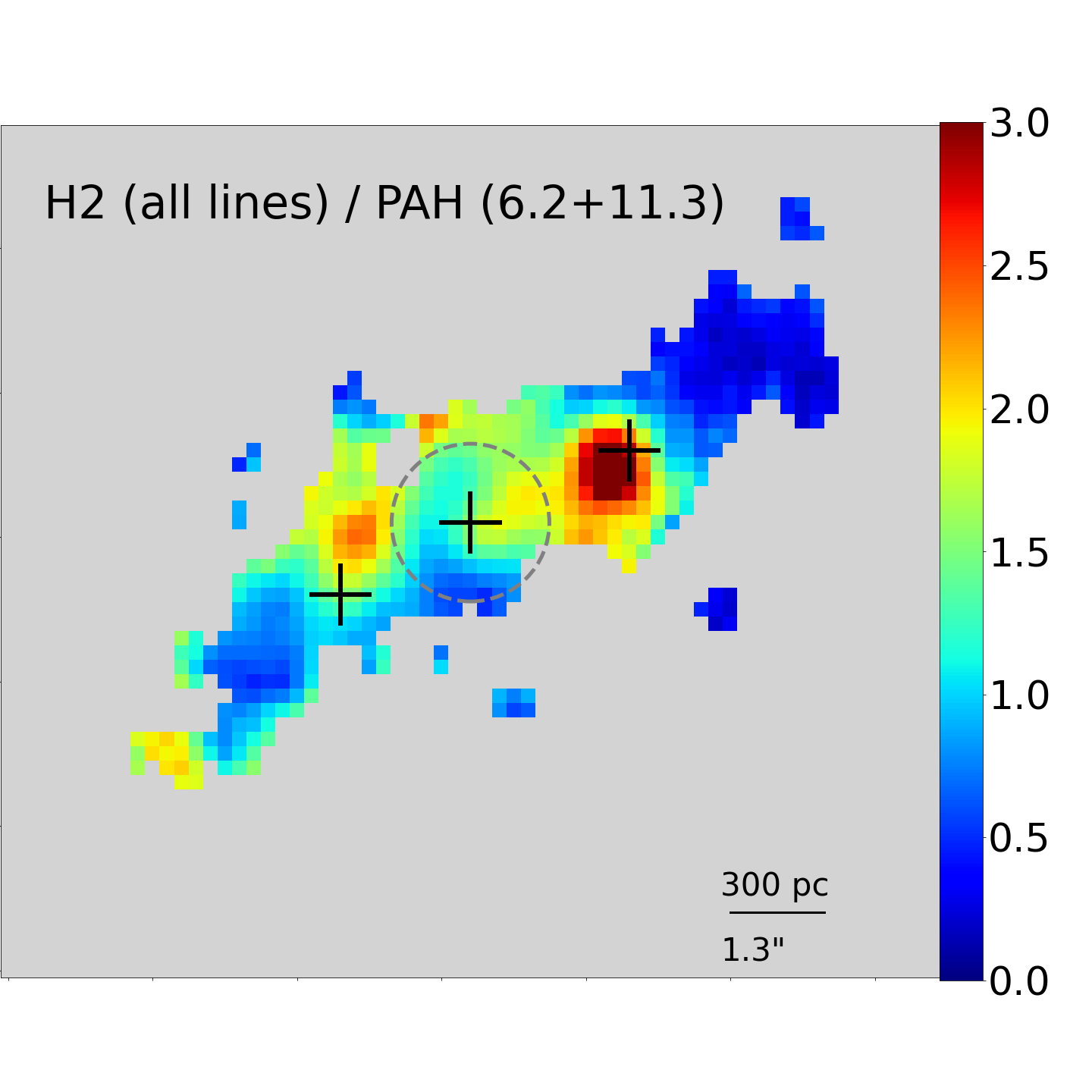}\\
\caption{ \htwo -to-PAH flux ratio, for the S(1), S(2), S(3) lines probing thermalized gas emission (left), and for all lines (right).
In both maps, the PAH emission comprises the flux in the two detected complexes at 6.2 $\mu$m and 11.3 $\mu$m.
\label{fig:H2_over_PAH}}
\end{figure}

\begin{figure}[ht!]
\centering
\includegraphics[width=0.35\textwidth]{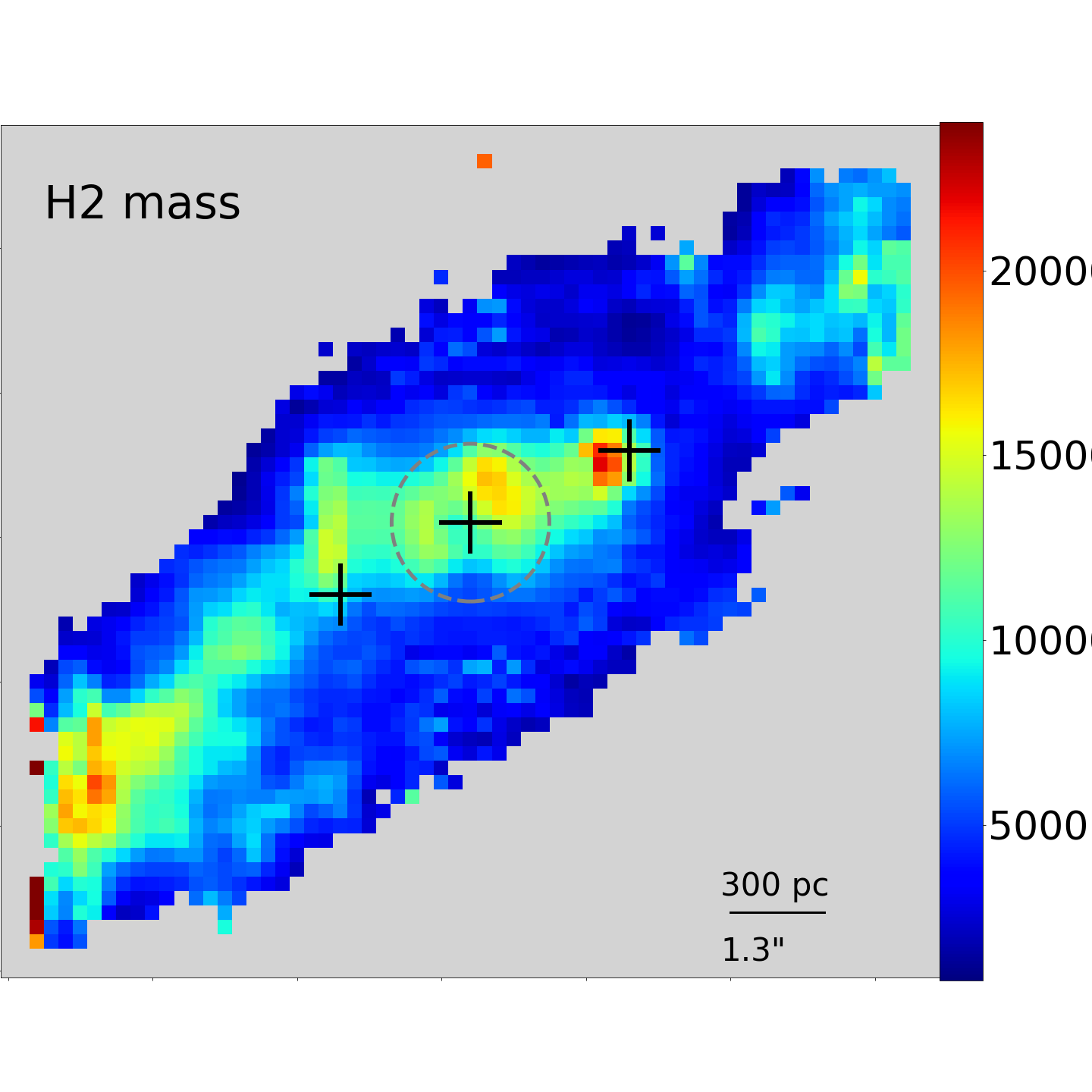}\\
\caption{ \htwo\ mass map, in units of solar masses, calculated for the coldest detected gas, probed by S(2) and S(1).
\label{fig:H2_mass}}
\end{figure}

The second reason has to do with the detection of bow-shock-shaped features and outflowing gas well outside of the 8 GHz and 17.8 GHz emission contours, 1-1.5 kpc away from the nucleus (while the presently detectable radio lobes reach only up to $\sim$500pc from the center). These features are compatible with a scenario in which the molecular gas has been dragged out by previous episodes of the radio jet, which was more extended or bright back then compared to the present-day radio jet. It is indeed frequent to observe recurrent radio activity in radio-loud objects (e.g. \citealt{owsianik98, kuzmicz17,shabala20}), and statistics on a large number of radio sources reveal a short duty cycle for radio jets, which could be explained by feedback-regulated accretion: in more than 10\% of the cases, there is evidence of radio lobe remnants, with a restarted activity in the center and with timescales between the two jet activities of about 1-10 Myr. In this scenario, the \htwo\ gas excitation has to have lasted longer, indicatively for timescales of a few Myrs, after the jet passed from a region or faded. Indeed, the heating of the molecular gas could be prolonged through shocks and the turbulence that they drive, with a dissipation timescale of the order of tens of Myrs, as the turbulent energy cascade and dissipation happens for many orders of magnitude in spatial scale (e.g. \citealt{ballesteros99,guillard09, appleton23}; see also \citealt{lebourlot99} for the cooling function of the \htwo).

In the center, i.e., in the region with detectable radio emission, the gas excitation can be elevated even in the case of shocks alone, because of superposition. Several shocks may be traveling in the ISM at different directions and velocities thanks to the jet scattering at various locations and thanks to the propagation of a slower moving, mass loaded gas cocoon that is linked to the heavier outflow \citep{morganti15}. If shocks significantly contribute to the heating, the gas is also likely to be more turbulent, showing broader line profiles in regions where the gas is hotter. The comparison to the velocity dispersion maps of several bright lines in Fig.~\ref{fig:moments} advocates in favour of this scenario.

The most appropriate scenario or combination of scenarios per region will be addressed in a a forthcoming paper using appropriate shock models that include cosmic ray heating, far UV photon heating, and (AGN or jet related) X-ray heating. The latter heating mechanism will also be taken into account, even though it was previously deemed to be rather weak at the position of the radio lobes: given the observed X-ray emission \citep{travascio21}, \citet{dasyra22} calculated that a heating rate of 10$^{-20}$ erg s$^{-1}$ cm$^{-3}$, which would matter for the CO excitation, could only be attained at the NW radio lobe if all clouds were near (e.g., within several parsecs of) a single X-ray source of $\sim$10$^{39}$ erg s$^{-1}$. We further add here that a substantial increase in the stellar far UV radiation would not be able to reproduce the CO emission either, so it is not a viable scenario by itself.

For the biconical outflow, seen in the west part of this region for 100$<$V$<$450 \kms , the same mechanisms may also be contributing to the gas excitation. The biconical outflow's excitation temperature also shows an increase in the S(3) and S(5) based T$_{ex}$ map, in which this outflow is clearly detected. There, T$_{ex}$ of 800$-$1100\,K are detected instead of 650-700\,K further out. These temperatures may be overestimated though, as the upper energy level (J=7) of S(5) could be non-thermalized: the critical density of this transition is 3.5$\times$10$^5$ cm$^{-3}$ \citep{lebourlot99} and fluorescence could then be contributing to the observed line emission. 

The contribution of lines from non-fully-thermalized states is instructive when studying the \htwo -to-PAH flux ratio.  A map of this ratio, when constructed from the thermal emission of the S(1), S(2), and S(3) lines  (Fig.~\ref{fig:H2_over_PAH}; left), reveals that the highest value is seen at the b1 bow shock. The elevated thermal \htwo\ over PAH emission found in numerous radio galaxies \citep{ogle10} can, thus, indeed be attributed to shocks. A map of the same ratio, when constructed from the total (thermal and fluorescent) emission of all \htwo\ lines (Fig.~\ref{fig:H2_over_PAH}; right) is instead brightest at c1, close to the north radio lobe. There, the maximum detected value (3.5) is 10 times greater than the average EELR value. This result could reflect the higher excitation of the molecular gas at the vicinity of the radio lobes, as presented above. Alternatively, an elevated \htwo -to-PAH ratio could signify the destruction of PAHs (e.g., \citealt{diamond-stanic10}), even though the PAH emission is bright along the jet axis (see also \citealt{ogle24}). As the \htwo -to-PAH ratio is an output of shock models (e.g., \citealt{kristensen23, villa24}), the PAH properties will be further discussed in our forthcoming paper.\\

\subsection{The molecular gas mass}
\label{disc:molecularmass}

We use the T$_{ex}$ map from S(1) and S(2) to determine the mass in each position of the galaxy, keeping in mind that this T$_{ex}$ indeed describes a (near) LTE temperature up to J=5. Combining this temperature with the S(1) luminosity map, we obtain the mass map shown in Fig.~\ref{fig:H2_mass}. The integrated mass in the FOV, M$_{\htwo}$ is  1.0$\times$10$^7$\msun . An ortho-to-para ratio of 3 is assumed for this calculation, which is reasonable for temperatures $>$200\,K (e.g., \citealt{rigopoulou02}). A more massive reservoir of \htwo\ gas does plausibly exist \citep{guillard12}, but its proper mass computation would require similar S(0) observations, as the mass measurement is highly sensitive to the T$_{ex}$ spatial distribution. For comparison, in a similar aperture, the cold \htwo\ mass from ALMA C0(1-0) data is 8.2($\pm$0.6)$\times$10$^8$\msun , when plugging the CO(1-0) flux (30.6$\pm$2.4 Jy\kms ; \citealt{dasyra22}) and a Galactic CO luminosity to \htwo\ mass factor $\alpha_{CO}$=4.6 \msun\ K$^{-1}$ km$^{-1}$ s pc$^{-2}$ \citep{bolatto13} into Equation 3 of \citet{solomon97}. For the cold gas, the mass in the outflow is too small (2$\times$10$^6$\msun ; \citealt{dasyra16}, to indicate that the bulk of the reservoir requires a sub-Galactic $\alpha_{CO}$.

Concerning individual regions of interest, the bow-shock shaped structure contains 8.8$\times$10$^5$\msun\ of warm molecular gas, while the mass around the NW radio lobe is similar, 7.8$\times$10$^5$ \msun . Of the latter mass, 60($\pm$14)\% is linked to the outflow in the \htwo, as indicated by spectral fitting with a double Gaussian. The computation of the mass flow rate of the accelerated gas requires knowledge of the deprojected velocity, the outflow geometry \citep{lutz19,veilleux20}, and the exact outflow starting point(s) $-$ thus the distance d traveled by the outflow(s). We performed an order of magnitude calculation for the NW radio-lobe outflow, assuming that it started locally, upon impact of the jet with clouds at the radio lobe. We used d in the range 60-250\,pc (i.e., the difference in the spatial offset between the CO(1–0) peak and the CO(4–3) peak in the ALMA data and the difference between the clump c2 and the NW lobe in the present JWST data), and an indicative velocity of 100\kms. Then, the mass flow rate within 1\arcsec\ from the NW radio lobe, computed as M$_{\htwo}$Vd$^{-1}$,  is in the range 0.1$-$1\msun yr$^{-1}$. These numbers also indicate an indicative outflow travel time, and thus a mininum jet age, of 0.6-2.4 Myrs. Similarly, within 1\arcsec\ from the SE radio lobe, the mass is 7.4$\times$10$^5$\msun , the outflow fraction is 40($\pm$10)\%, and the contribution to the total mass flow rate is just a little lower than that of the NW lobe. Still, these numbers are low enough that the mass flow rate is dominated by the CO and HI emitting gas \citep{morganti07,morganti15,dasyra16}, even if some \htwo\ were to be unaccounted for due to the lack of S(0) data. \\

\section{Conclusions} 
\label{sec:conclusions}

We obtained JWST MIRI data of the nucleus of the nearby galaxy IC5063, in which a jet is known to interact with dense gas clouds and drive multi-phase outflows, changing the gas excitation. Our new JWST data provided IFU datacubes of the inner $\sim$3$\times$2kpc$^2$ of the galaxy (in the common FOV of all observations), and revealed the following findings.
\begin{itemize}
    \item An unresolved source at the galaxy's nucleus emits a bright MIR continuum, radially limited to 50pc, that can be attributed to the torus with potential contribution from the radio core.
    \item Thirty nine spectral lines are detected on top of the continuum. Of those, the best tracers of the gaseous disk are the low-J \htwo\ lines, (0-0) S(1) and S(2). Instead, the ionized gas traces closely the jet trail and the optical EELR. Higher ionization potential species tend to have a more nucleated emission. Given its IP and the spatial resolution of MIRI at its wavelength, \mgv\ is the best tracer of the jet trail.
    \item Outflows are identified in more than ten discrete locations of IC5063. These include the vicinity of the two radio lobes, the nucleus, and a biconical structure perpendicular to the jet trail, which host previously known outflows. Additionally, new outflows are detected at points near or intermediate to the above, at the outer edges of the EELR, in the interior of the EELR, and in an ionized gas bubble. Overall, outflows and turbulent kinematics are seen in the central $\sim$3.5 kpc$^2$. The outflow velocities reach and occasionally exceed 1000 \kms\ for the ionized gas. In the bright \siv\ line, within $\sim$600\,pc of the nucleus, the outflow velocities often exceed the local escape velocity. For the molecular gas, the observed velocities are lower, typically not exceeding the escape velocity. 
    \item For both gas phases, greater outflow velocities are observed for higher ionization/excitation lines near the radio lobes. This result indicates that the outflows are stratified, with the higher ionization/excitation gas being in closer proximity to the power source (i.e., the jet).    
    \item Newly detected bow shocks are seen in \htwo\ lines in regions without significant radio emission, further away from the nucleus than the radio lobes. Their presence suggests that a jet has propagated beyond what radio images indicate. Located near the edge of the SE EELR, these bow shocks indicate that the jet got diverted or scattered at the radio lobe near the SE EELR base, that it propagated along the EELR edge, and that it altered the gas distribution, kinematics, and ionization there. Outflowing gas is detected all along this EELR edge, for $\sim$1 kpc, including at the bow shock locations. Shock-driven turbulence is the most likely warm \htwo\ excitation mechanism in the bow shocks, as turbulent dissipation happens at much slower timescales than the passage of a relativistic jet.
    \item The most highly excited warm \htwo\ gas can be found where the most severe jet-cloud interactions take place: at the radio lobes, in the biconical outflow, and potentially at the nucleus. The entire region with radio emission shows a (step-function) increase of at least 100K in the excitation temperature compared to regions without radio emission, such as the EELR interior. More jet-related gas excitation mechanisms must therefore operate in that region, including cosmic rays and numerous shocks traveling at different directions/velocities. 
    \item The mass of the warm \htwo\ ($>$200K) is 1.0$\times$10$^7$\msun, as computed from the S(1) line emission map and the T$_{ex}$ map based on S(1) and S(2) . This value is nearly two orders of magnitude below that of the cold \htwo\ gas probed by CO lines. The molecular gas masses of both the disk and the outflow are dominated by the cold, CO-emitting gas component. 
\end{itemize}

\begin{acknowledgments}

{\noindent Acknowledgements: }This work is based on observations made with the NASA/ESA/CSA JWST. The data were obtained from the Mikulski Archive for Space Telescopes at the Space Telescope Science Institute. KMD would like to thank several members of the JWST staff, namely D. Law for a lengthy and detailed communication on the flux calibration, as well as the ESA/STScI team members M. Garcia-Marin, S. Kendrew, and K. Larson for pertinent discussions. US-based authors acknowledge support by NASA under the JWST data analysis grant JWST-GO-02004.002-A. GFP acknowledges support by the European Research Council advanced grant “M2FINDERS - Mapping Magnetic Fields with INterferometry Down to Event hoRizon Scales” (Grant No. 101018682). TGB acknowledges support from the Leading Innovation and Entrepreneurship Team of Zhejiang Province of China (Grant No. 2023R01008). JAFO acknowledges financial support by the Spanish Ministry of Science and Innovation (MCIN/AEI/10.13039/501100011033), by ``ERDF A way of making Europe'' and by ``European Union NextGenerationEU/PRTR'' through the grants PID2021-124918NB-C44 and CNS2023-145339; MCIN and the European Union -- NextGenerationEU through the Recovery and Resilience Facility project ICTS-MRR-2021-03-CEFCA. TGB acknowledges support from the Leading Innovation and Entrepreneurship Team of Zhejiang Province of China (Grant No. 2023R01008). 

\end{acknowledgments}

{}

\appendix
\label{Appendix}

\section{Spectral line and continuum maps}

\subsection{ Spectral line emission  }

\begin{figure}[ht!]
\centering
\includegraphics[width=\textwidth, trim=0cm 2cm 0cm 4cm, clip]{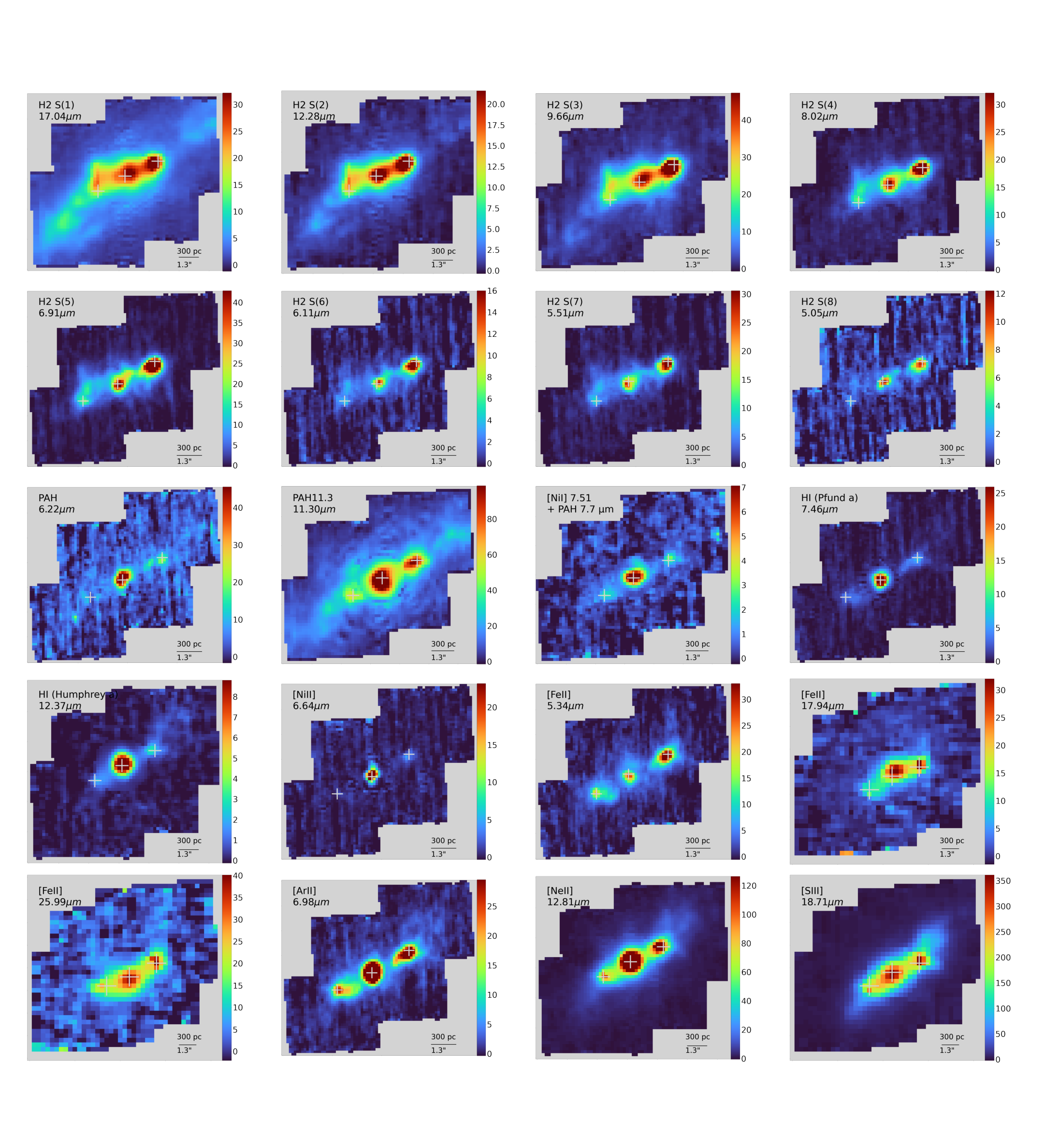}
\caption{ Detected gas emission, indicatively grouped by gas phase, in units of 10$^{-20}$ W m$^{-2}$. Ionized gas lines are presented in an order of increasing ionization potential.  Crosses mark the location of the nucleus and the EELR bases, as derived from the maxima of all ionized gas lines in our data, at rest-frame velocity.
\label{fig:all_lines}}
\end{figure}

The emission of all detected spectral lines, integrated over all wavelengths showing relevant emission, is shown in ~\ref{fig:all_lines}. All lines show extended emission. Typically, the emission becomes more nucleated with increasing IP.

\begin{figure}[ht!]
\centering
\includegraphics[width=\textwidth, trim=0cm 2cm 0cm 4cm, clip]{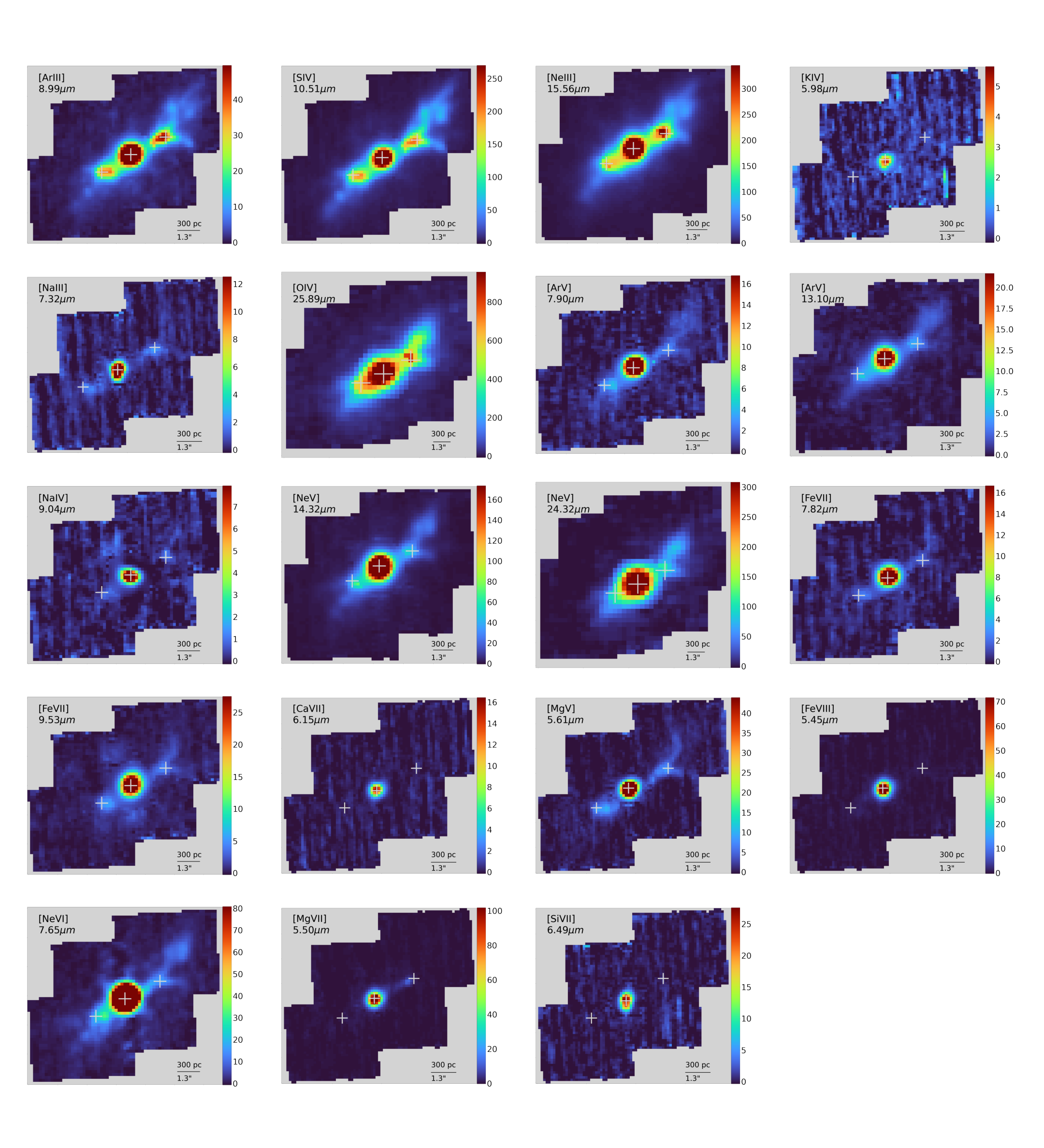}
{ \newline Figure ~\ref{fig:all_lines}:   Continued. 
}
\end{figure}

\subsection{ Continuum emission}
\label{Appendix:cont}
\begin{figure}[ht!]
\centering
\includegraphics[width=0.33\textwidth]{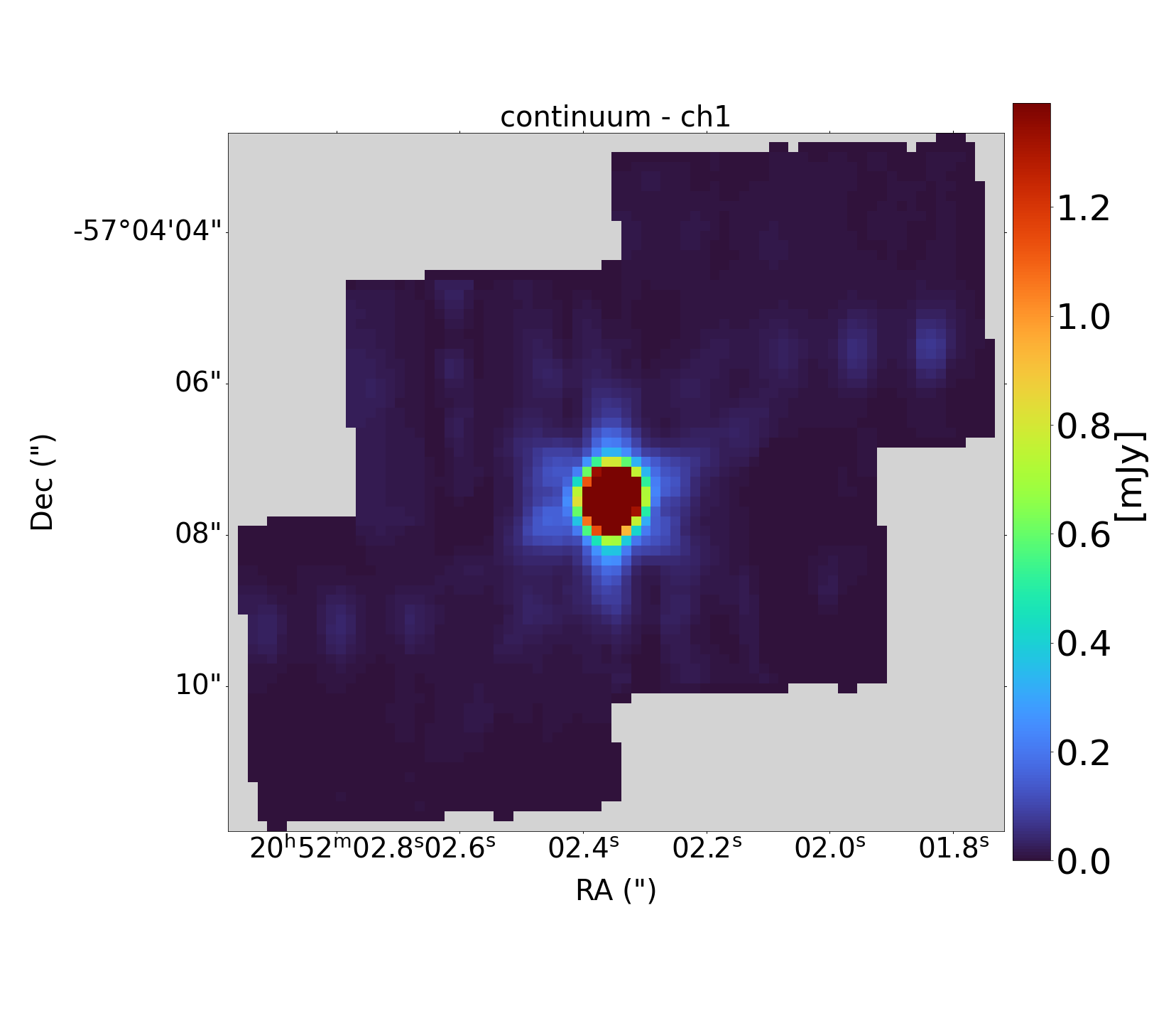}
\includegraphics[width=0.33\textwidth]{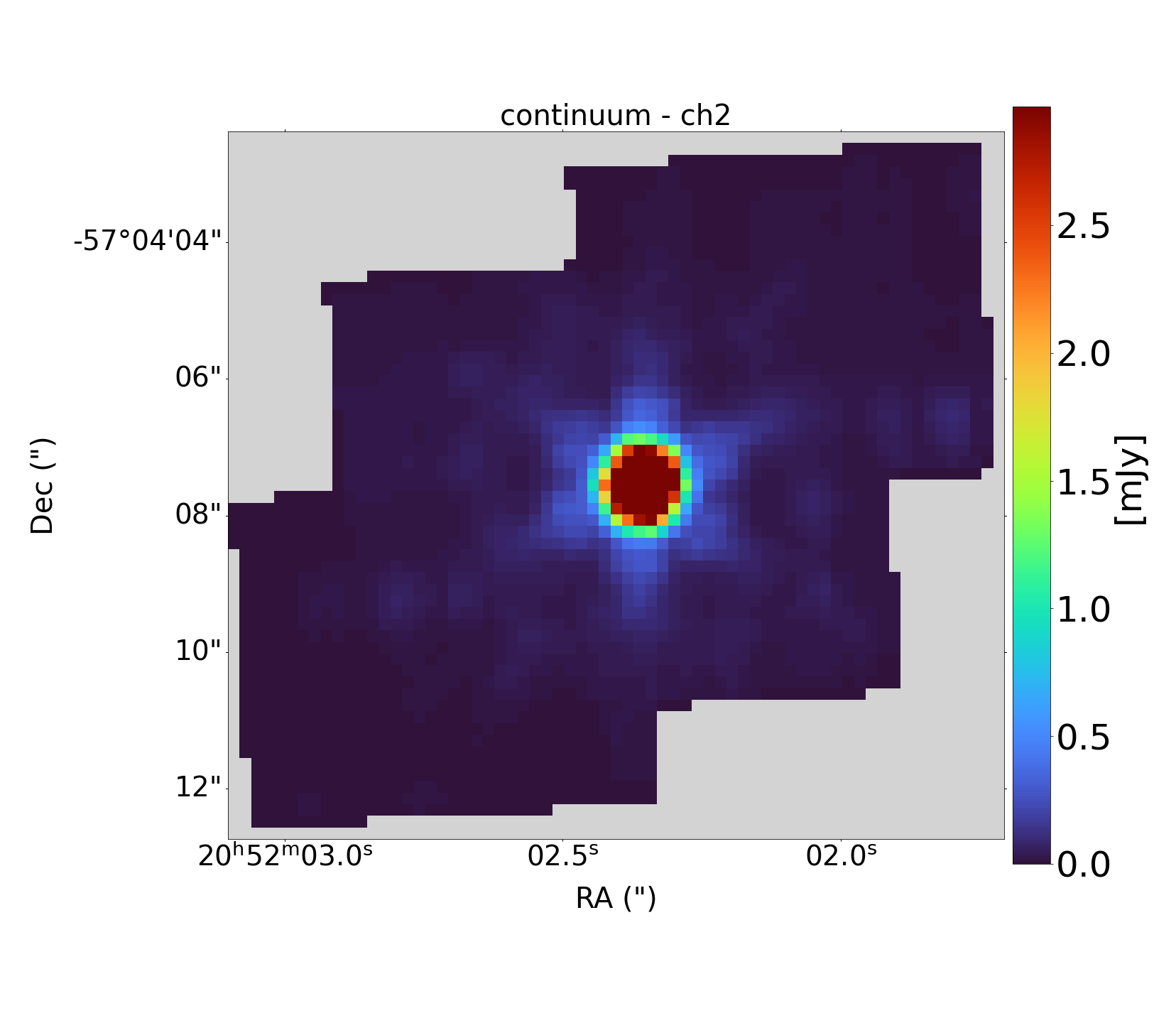}\\
\includegraphics[width=0.33\textwidth]{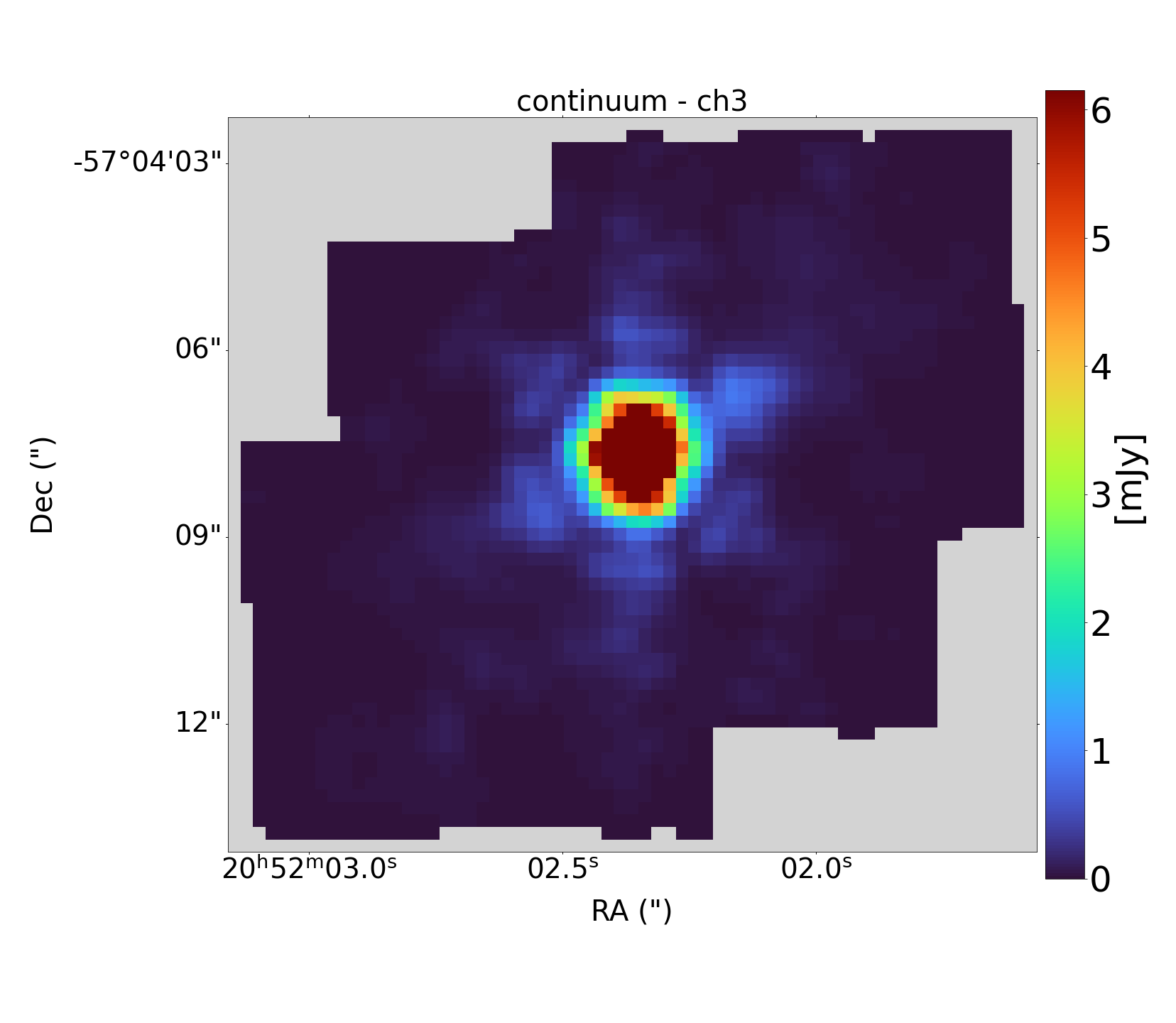}
\includegraphics[width=0.33\textwidth]{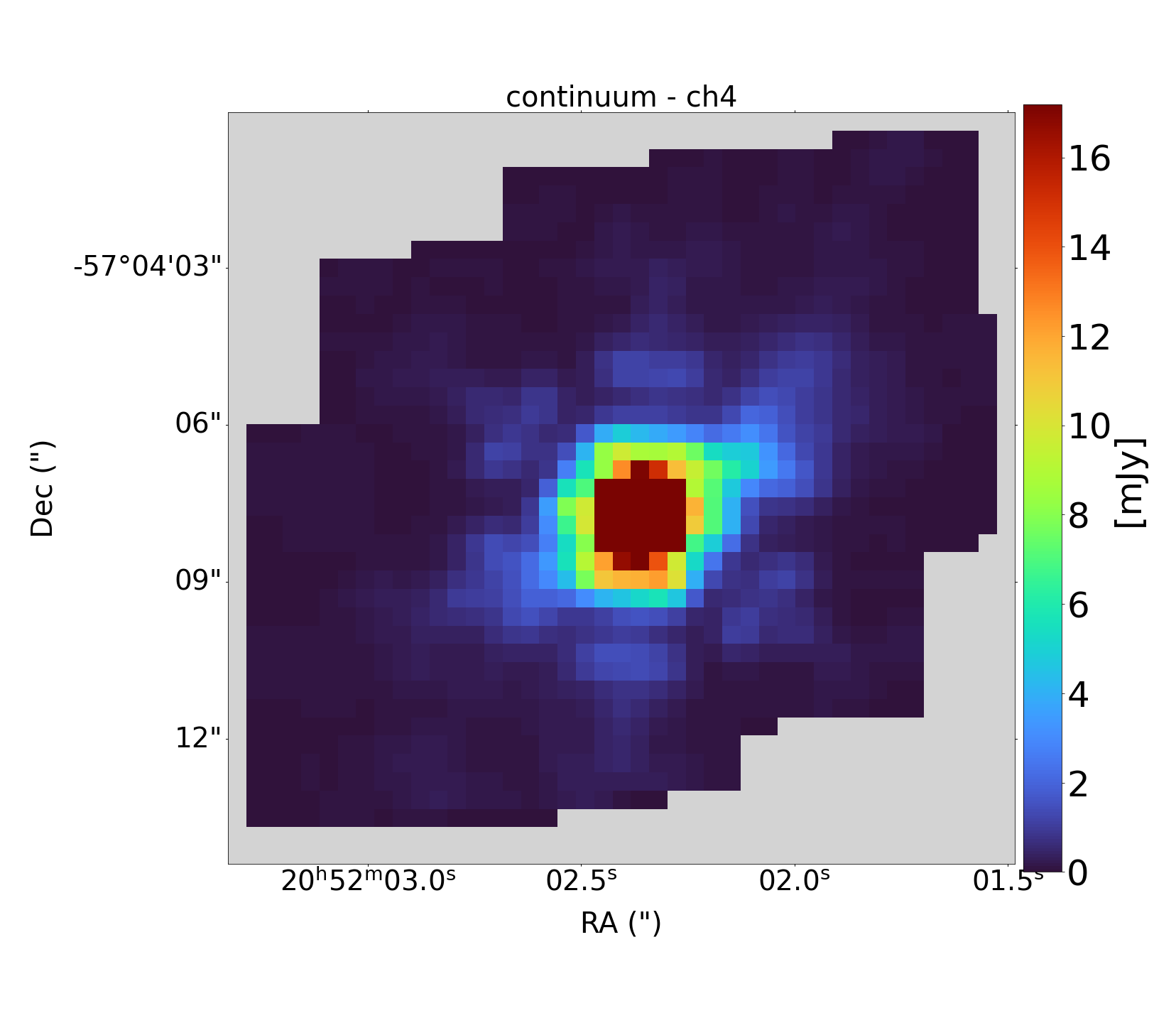}\\
\caption{ Continuum images in the medium band of each channel (i.e., ch1 to ch4 from left to right). 
\label{fig:cont}}
\end{figure}

The image of the continuum in each of the four MIRI channels, representing the instrument's PSF, is shown in Fig.~\ref{fig:cont}. The figure reveals the domination of the PSF over any other continuum source in the FOV. 

\section{Notes on the instrument's behaviour in the presence of bright line emission}

\subsection{Data showing the cruciform artifact}
\label{Appendix:cruciform}

\begin{figure}[ht!]
\centering
\hspace*{-1.5cm}
\includegraphics[width=1.15\textwidth]{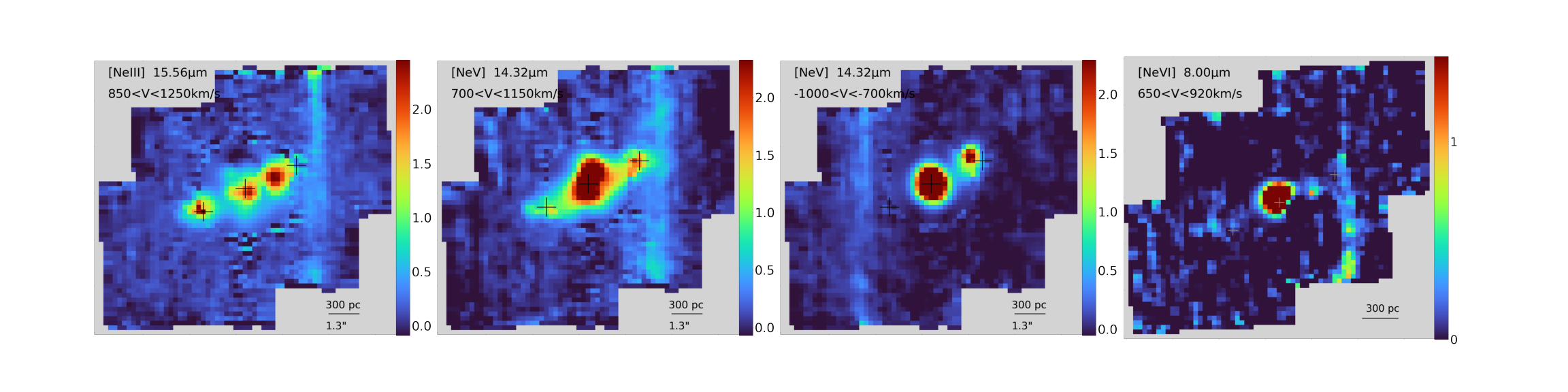  } 
\caption{Emission due to the cruciform artifact.
\label{fig:crux_emission}}
\end{figure}

\begin{figure}[ht!]
\centering

\includegraphics[width=0.68\textwidth]{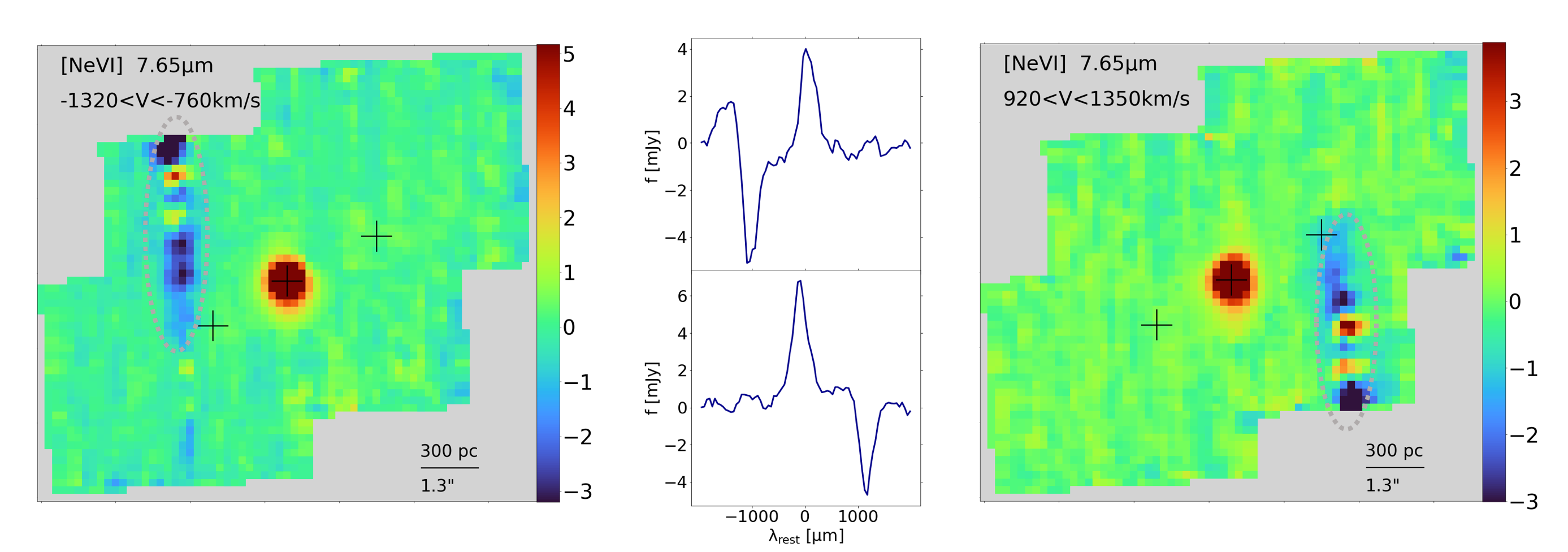}
\includegraphics[width=0.68\textwidth]{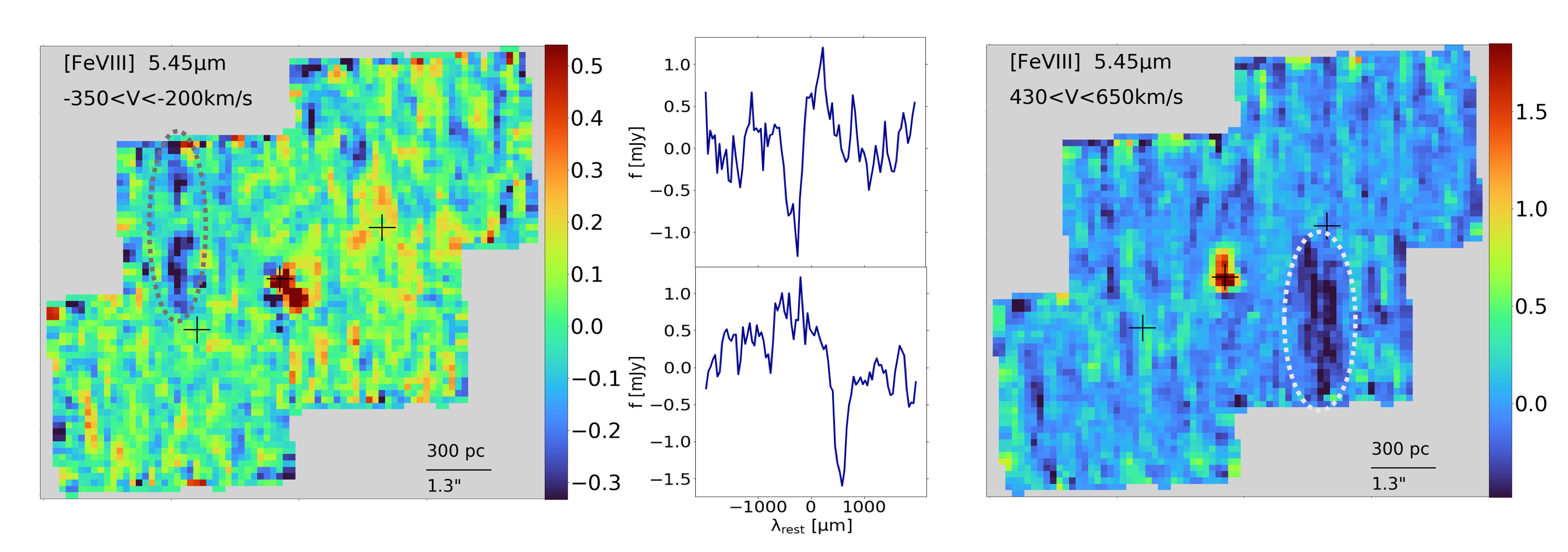}
\includegraphics[width=0.68\textwidth]{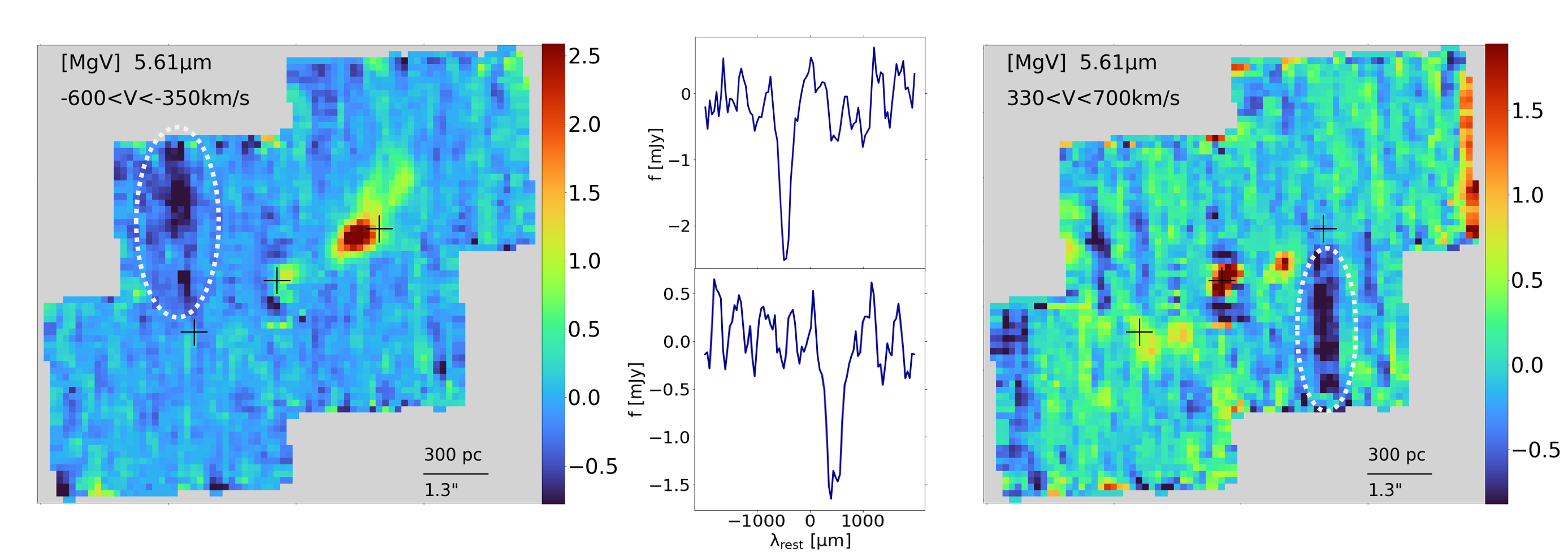}

\caption{ Absorption due to the cruciform artifact.
\label{fig:crux_absorption}}
\end{figure}

Besides the PSF residuals that are related to a bright continuum source, there is an artifact that can affect any part of the detectors that suffers from the leakage of photons from a bright emission line along the x axis of the raw data. The issue is reported and depicted for the imager in the JWST user documentation webpage\footnote{https://jwst-docs.stsci.edu/known-issues-with-jwst-data/miri-known-issues\#MIRIKnownIssues-cruciformCruciformartifact}. It is the so-called cruciform artifact. It originates from internal scattering of the light within the MIRI detectors, in particular at lower wavelengths, which have a $<$27\% quantum efficiency. Photons that are neither absorbed nor able to escape (in particular for outer PSF regions) are diffracted by the lattice and cause straight-light patterns \citep{gaspar21,argyriou23}. This was occasionally clearly seen in our data, as stripes oriented north-south (e.g., in the reshifted \neiii\ and \nev\ emission shown in Fig.~\ref{fig:crux_emission}). The cruciform was more easily seen for bright emission lines. Bright lines contaminate neighboring detector slices at a constant y-axis pixel. But both because of the detector distortion and also because neighboring detector slices do not correspond to neighboring sky positions, the contamination appears at different wavelengths from the emission line peak wavelength and in various positions in the reconstructed cubes' FOV. Moreover, because of diffraction, the stray light can either be seen as a stripe, or as an Airy diffraction pattern (see the \nevi\ emission in Fig.~\ref{fig:crux_emission}). The pipeline, by default, automatically subtracts a (straight line emission) model in channels 1 and 2, in an attempt to remove this stray light. However, because of the detector distortion, signal may be oversubtracted from some regions and undersubtracted in other regions. This leads to regions of images that appear to have cavities (e.g., \neiii\ in Fig.~\ref{fig:crux_emission}) and/or strong absorption (as in all panels of Fig.~\ref{fig:crux_absorption}), that can contain emission blobs within the absorption area. This can be confused for real absorption in case there is a bright, warm background emission source such as a jet's synchrotron in the background. Or it could be confused for a P-Cygni or inverse P-cygni profile. In our case, there are winds \citep{dasyra15}, but there is no detected synchrotron emission in the regions where the absorption is seen. (Note that the images and spectra shown in Fig.~\ref{fig:crux_absorption} are continuum subtracted.) Therefore, the absorption would have to be tracing outflows of cold, yet highly ionized gas that is absorbing the emission of hot dust in the background - a rather unusual configuration. However, we report the presence of the artifact for demonstration purposes.

\subsection{Data with signal potentially linked to the cruciform artifact}
\label{Appendix:potentia_outflows}

\begin{figure}[ht!]
\centering
\hspace*{-1.5cm}
\includegraphics[width=1.15\textwidth]{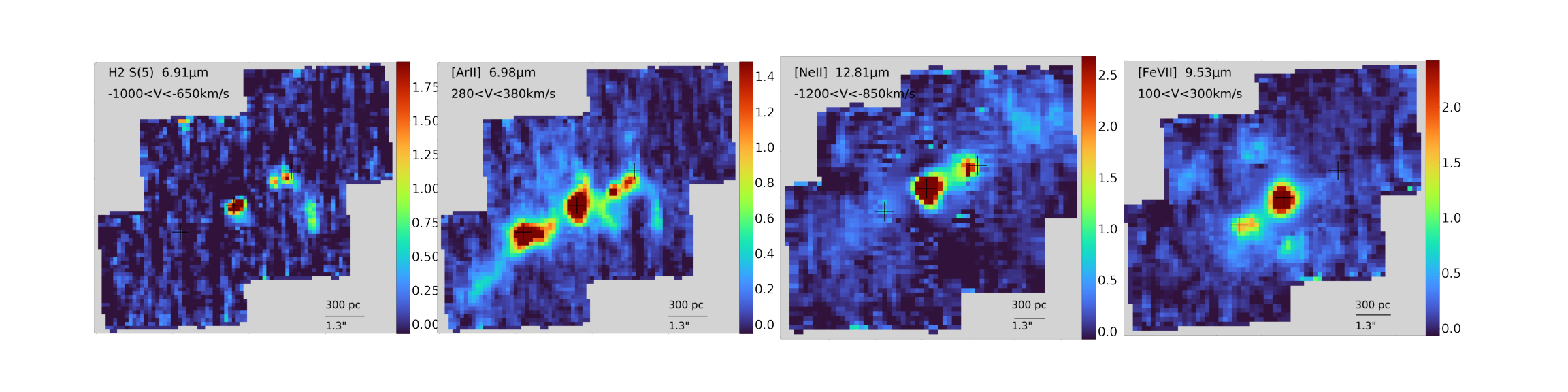  } 
\caption{Emission that could be associated with regions of interest, but that cannot be confirmed to be real due to the cruciform artifact and PSF residuals.
\label{fig:potential_emission}}
\end{figure}

Because of this artifact and PSF residuals, we occasionally see signal at low or high velocities that cannot be certainly determined to be true. Examples of regions (except the nucleus, radio lobes, and already reported knots) where more outflows could potentially exist are shown in Fig.~\ref{fig:potential_emission}. The presence of outflows in these regions needs to be tested with other data or MIRI data of different setup. A most striking feature is a filament that starts from the NW radio lobe, then tilts and continues in a north-south orientation. It is presented in the selected \arii\ and \htwo\ S(2) images of Fig.~\ref{fig:potential_emission}, but it is present in a plethora of lines.

\end{document}